\documentclass{article}
\usepackage[utf8]{inputenc}
\pdfoutput=1

\usepackage{enumerate}

\usepackage{appendix}
\usepackage{mathrsfs}
\usepackage{amsmath}
\usepackage{amssymb}
\usepackage{float}
\usepackage{graphicx}
\usepackage{multirow}
\usepackage{xcolor}
\usepackage{bm}
\usepackage[normalem]{ulem}
\usepackage[autostyle]{csquotes} 

\usepackage{booktabs}
\usepackage{color}
\usepackage{jcappub}

\def\vk{{\bf k}}
\def\fnl{f_{\rm NL}}
\def\dh{\delta_{\rm h}}
\def\dm{\delta_{\rm m}}

\title{Taming assembly bias for primordial non-Gaussianity}
\author[1]{Emanuele Fondi,}
\author[1,2]{Licia Verde,}
\author[3,4]{Francisco Villaescusa-Navarro,}
\author[5,6,7]{Marco Baldi,}
\author[8,9,4]{William R.~Coulton,}
\author[10]{Gabriel Jung,}
\author[11]{Dionysios Karagiannis,}
\author[12,13]{Michele Liguori,}
\author[12,13]{Andrea Ravenni,}
\author[14,3]{Benjamin D.~Wandelt}

\affiliation[1]{ICC, University of Barcelona, IEEC-UB, Mart\' i i Franqu\` es, 1, E08028
Barcelona, Spain}
\affiliation[2]{ICREA, Pg. Lluis Companys 23, Barcelona, 08010, Spain} 
\affiliation[3]{Center for Computational Astrophysics, 162 5th Avenue, New York, NY, 10010, USA}
\affiliation[4]{Department of Astrophysical Sciences, Princeton University, 4 Ivy Lane, Princeton, 
NJ 08544 USA}
\affiliation[5]{Dipartimento di Fisica e Astronomia, Alma Mater Studiorum - University of Bologna, Via Piero Gobetti 93/2, 40129 Bologna BO, Italy}
\affiliation[6]{INAF - Osservatorio Astronomico di Bologna, Via Piero Gobetti 93/3, 40129 Bologna BO, Italy}
\affiliation[7]{INFN - Istituto Nazionale di Fisica Nucleare, Sezione di Bologna, Viale Berti Pichat 6/2, 40127 Bologna BO, Italy}
\affiliation[8]{ Kavli Institute for Cosmology, Madingley Road, Cambridge, CB3 0HA, UK}
\affiliation[9]{DAMTP, Centre for Mathematical Sciences, Wilberforce Road, Cambridge CB3 0WA, UK}
\affiliation[10]{Universit\'{e} Paris-Saclay, CNRS, Institut d’Astrophysique Spatiale, 91405, Orsay, France}
\affiliation[11]{Department of Physics \& Astronomy, University of the Western Cape, Cape Town 7535, South Africa}
\affiliation[12]{INFN, Sezione di Padova, via Marzolo 8, I-35131, Padova, Italy}
\affiliation[13]{Dipartimento di Fisica e Astronomia “G. Galilei”, Università degli Studi di Padova, via Marzolo 8, I-35131, Padova, Italy}
\affiliation[14]{Sorbonne Universit\'{e}, CNRS, UMR 7095, Institut d'Astrophysique de Paris, 98 bis bd Arago, 75014 Paris, France}

\emailAdd{emanuelefondi@icc.ub.edu, liciaverde@icc.ub.edu}

\abstract{Primordial non-Gaussianity of the local type  induces a  strong scale-dependent bias on the clustering of halos in the late-time Universe. This signature  is particularly promising to provide constraints on the non-Gaussianity parameter $f_{\rm NL}$  from galaxy surveys, as the bias amplitude grows with scale and becomes important on large, linear  scales.  However, there is a  well-known degeneracy between the real prize, the $f_{\rm NL}$ parameter, and the (non-Gaussian) assembly bias  i.e., the halo formation history-dependent  contribution to the amplitude of the signal, which could seriously compromise the ability of large-scale structure surveys to constrain $f_{\rm NL}$. We show how the assembly bias can be modeled and constrained, thus almost completely  recovering the power of galaxy surveys to competitively constrain primordial non-Gaussianity. In particular, studying hydrodynamical simulations, we find that a proxy for the halo properties that determine assembly bias can be constructed from photometric properties of galaxies. Using a prior on the  assembly bias guided  by this proxy degrades the statistical errors on $\fnl$ only mildly compared to an ideal case where the assembly bias is perfectly known. The systematic error on $\fnl$ that the proxy induces  can be  safely kept under control.} 
\begin{document}
\maketitle
\flushbottom
\section{Introduction}
Inflation represents the standard paradigm for the description of the early evolution of our Universe. In its simplest form, it is determined by a single slow-roll scalar field and produces an almost Gaussian \cite{maldacena_non-gaussian_2003} distribution of primordial perturbations, which evolve gravitationally to form structure in the late Universe. Although current constraints from Cosmic Microwave Background (CMB) \cite{planck_collaboration_planck_2019} and large-scale structure (LSS) \cite{mueller_clustering_2021} observables confirm the Gaussian scenario, the enhanced sensitivity of ongoing \cite{desi_collaboration_desi_2016,amendola_cosmology_2018} and forthcoming \cite{dore_cosmology_2015,schlegel_astro2020_2019} LSS experiments may have the potential to reveal the non-Gaussianity of the initial conditions, predicted by alternative inflationary models \cite{giannantonio_constraining_2012,alvarez_testing_2014,de_putter_designing_2017,karagiannis_constraining_2018,ferraro_snowmass2021_2022,achucarro_inflation_2022}.

In particular, the presence of multiple fields during inflation would introduce non-linearities in the primordial Bardeen potential, generating what is known as Primordial non-Gaussianity (PNG) of the \textit{local} type \cite{gangui_three--point_1994, 2000PhRvD..61f3504W, bernardeau_non-gaussianity_2002}. Local PNG can be expressed in its simplest form as a quadratic correction with amplitude $\fnl$ \cite{verde_large-scale_2000,komatsu_acoustic_2001}:
\begin{equation}
\Phi({\bf x})=\phi({\bf x})+\fnl\left(\phi({\bf x})^2-\left<\phi({\bf x})^2\right>\right)
\label{eq:fnl}
\end{equation}
where $\Phi({\bf x})$ is the Bardeen potential and $\phi({\bf x})$ is an auxiliary Gaussian random field.

The primordial bispectrum generated by local PNG couples small scale density perturbations with large scale potential modes, affecting dark matter halo clustering at late times. As a result, the halo bias takes a scale dependence on large scales, with an amplitude given by $b_{\phi}\fnl$, where $b_{\phi}$ is the local PNG bias parameter \cite{dalal_imprints_2008,matarrese_effect_2008,slosar_constraints_2008,afshordi_primordial_2008,mcdonald_primordial_2008}.
Detecting a nonzero $b_{\phi}\fnl$ would be sufficient to rule out the standard single-field, slow-roll inflationary model \cite{creminelli_single_2004}. However, in order to constrain the value of $\fnl$ and shed light on the physics of inflation, discriminating between different models, a prior knowledge of $b_{\phi}$ is required.

Conventionally, predictions for $b_{\phi}$ rely on the universality relation, derived under the assumption of a halo mass function (HMF) which is \textit{universal}, i.e., dependent on halo mass and redshift only through a single variable \cite{bond_excursion_1991,press_formation_1974,sheth_large_1999}. However, a number of recent results from N-body simulations highlighted the breakdown of this relation, due to assembly bias in $b_{\phi}$, i.e., its dependence on properties beyond the halo mass. Specifically, at fixed mass,  halos with different concentration have significantly different $b_{\phi}$ \cite{lazeyras_assembly_2022,barreira_towards_2023,sullivan_learning_2023}. Similar conclusions have been reported for samples of galaxies selected by their stellar mass, color and other galaxy properties \cite{barreira_impact_2020,barreira_galaxy_2020,barreira_can_2022}.

Recognizing these departures from universality paved the way for further studies in various directions. In particular, the theoretical uncertainties called for new efforts to provide better priors on $b_{\phi}$ with the aid of machine learning techniques \cite{sullivan_learning_2023,lucie-smith_halo_2023}, or to break the $b_{\phi}\fnl$ degeneracy introducing new summary statistics \cite{jung_quijote-png_2023}. Furthermore, since samples with different $b_{\phi}$ represent an opportunity to employ the multi-tracer technique \cite{seljak_measuring_2009} and provide better $\fnl$ constraints, some analyses focused on optimizing sample selection strategies \cite{sullivan_learning_2023,barreira_towards_2023,karagiannis_multi-tracer_2023}.

Although recently highlighted in many works, these issues were already pointed out by the seminal papers in the field \cite{slosar_constraints_2008,seljak_measuring_2009,slosar_optimal_2009,reid_non-gaussian_2010}. In particular, the attention was focused on how details of the halo formation history affect the value of $b_{\phi}$. Specifically, Ref. \cite{slosar_constraints_2008} showed that at fixed mass, halos which assembled their mass recently have a $b_{\phi}$ lower than predicted by the universality relation. This result was generalized by Ref. \cite{reid_non-gaussian_2010}, where the authors effectively found an analytical relation between $b_{\phi}$ and halo formation time, by employing the extended Press-Schechter (ePS) formalism \cite{lacey_merger_1993,lacey_merger_1994,bosch_universal_2001}. In the same way as the Press-Schechter formalism can be used to compute a universal HMF, its \textit{extension} gives the conditional mass function (CMF) that describes the halo mass accretion history. The CMF can be used to estimate the departure of $b_{\phi}$ from the universality relation.

In this paper we use different suites of N-body simulations \cite{villaescusa-navarro_quijote_2020,coulton_quijote_2022,coulton_quijote-png_2022,jung_quijote-png_2022} to test the validity of the analytical ePS prediction, across various halo mass ranges and redshifts. To take into account inconsistencies, we suggest a 1-parameter extension of the ePS, calibrated on simulations. We then resort to hydrodynamical simulations \cite{nelson_illustristng_2021,nelson2018first,villaescusa-navarro_camels_2021} which also model galaxy formation to further connect the modeling of $b_\phi$ and assembly bias to observational properties of galaxies. In particular, we study photometric properties of simulated galaxies  to  find an observational proxy for halo assembly bias and assess its robustness to changes in cosmological and astrophysical parameters. Finally, we report $\fnl$ error forecasts for single and multi-tracer analyses, providing insights on optimal galaxy sample selection strategies.

The rest of the paper is structured as follows. In Section~\ref{sec:thbackground}  we review the theoretical background of the non-Gaussian halo bias and the ePS prediction for $b_{\phi}$. In Section~\ref{sec:ePStest} we test the ePS predictions on N-body simulations and calibrate a simple extension of the analytical result, which shows a remarkable good fit across different halo mass and redshift ranges. In Section~\ref{sec:proxy} we consider different galaxy samples and identify the optimal proxy for halo assembly bias from combinations of photometric galaxy colors. The robustness of this method is assessed in Section~\ref{sec:camels} by using a different suite of simulations, which consider a broad range in the astrophysical parameters. Focusing on ELG-like and LRG-like simulated samples, we report Fisher forecasts for the precision of the $\fnl$ measurement from single and multi-tracer analyses in Section~\ref{sec:forecasts}. We finally summarize our main results and conclude in Section~\ref{sec:conclusions}.

\section{Theoretical background}
\label{sec:thbackground}
In this section we summarize the relevant results from the literature. While no novel results are presented, the section serves to lay out useful equations and formulae and to provide a unified notation.

Predictions for the distribution and clustering of dark matter halos in the presence of PNG can be obtained using conceptually distinct but mutually consistent approaches \cite{dalal_imprints_2008,matarrese_effect_2008,slosar_constraints_2008,afshordi_primordial_2008}. In the peak-background split formalism \cite{sheth_large_1999}, the auxiliary Gaussian field $\phi$ in Eq.~\ref{eq:fnl} is decomposed into statistically independent  short- and long- wavelength modes $\phi=\phi_l+\phi_s$.

The gravitational potential fluctuations described by Eq.~\ref{eq:fnl} are translated into (linear) matter density fluctuations at redshift $z$ through the Poisson equation
\begin{equation}
\dm(\vk, z)\equiv{\cal M}(k,z)\Phi(\vk)=\frac{2c^2k^2T(k)}{3\Omega_mH_0^2}D(z)\Phi(\vk)
\end{equation}
where $D(z)$ denotes the linear growth rate of perturbations normalized to be $(1+z)^{-1}$ during matter-domination and $T(k)$ being the matter transfer function normalized to unity on large scales.

In the absence of PNG, small- and large-scale linear density fluctuations would evolve independently under the Poisson equation, while in the presence of a non-zero $\fnl$ in Eq.~\ref{eq:fnl}, the short wavelengths in the density field become:
 \begin{equation}
\delta_s={\cal M}(k,z)[(1+2\fnl\phi_l)\phi_s+\fnl\phi_s^2].
\label{eq:ds}
 \end{equation}
Thus small scale density fluctuations are modulated by large scale potential modes, with a coupling given by $\fnl$. In turn, the number of collapsed halos will be affected, determined both by density and potential fluctuations. Writing the halo overdensity $\dh$ as a linear bias expansion, we can therefore express it in terms of the two fields $\delta_m$ and $\Phi$ as \cite{mcdonald_primordial_2008}
\begin{eqnarray}
\dh(\vk|M,z) &=& b_1(M,z) \dm(\vk,z) + b_{\phi}(M,z) \fnl \Phi(\vk) + \epsilon(\vk,z) \nonumber \\ 
&=& \left[b_1(M,z) + \frac{b_{\phi}(M,z) \fnl}{{\cal M}(k,z)}\right]\dm(\vk,z)+ \epsilon(\vk,z)
\label{eq:dh_expansion}
\end{eqnarray}
where $\epsilon$ denotes a shot noise (or stochastic) term and $b_1$ and $b_{\phi}$ are bias parameters. The first bias term in the bracket is scale-independent, while the second one, which depends on $k$ via ${\cal M}(k,z)$ represents the so-called \textit{scale-dependent bias}. Both bias parameters can be related to the halo mass function $n(M,z)$, i.e. the number density of halos in a mass interval $dM$ around  mass $M$ and  redshift interval $dz$ around redshift $z$.

In the Gaussian case, the  linear Lagrangian bias reduces to the scale-independent term:
\begin{equation}
b_1^{G}(M,z)=\frac{\partial \ln n^G(M,z)}{\partial \delta_l} =-\frac{\partial \ln n^G(M,z)}{\partial \delta_c}
\end{equation}
where $n^G$ denotes the Gaussian halo mass function. The second equality arises because the effect of modulating the density field by a long wavelength mode $\delta_l$ in some patch of the Universe can be seen as an additive change in the  critical density for collapse $\delta_c$ in that region \cite{cole_biased_1989}.

In the presence of non-Gaussianity there are two main corrections.
The first one regards the scale-independent bias parameter $b_1$ \cite{desjacques_scale-dependent_2009,Wagner:2011wx}:
\begin{equation}
    b_1(\fnl,M,z)=b_1^G(M,z)+\Delta b_I(\fnl,M,z)=\frac{\partial \ln n^{NG}}{\partial \delta_l}=-\frac{\partial \ln n^G}{\partial\delta_c}-\frac{\partial\ln R^{NG}}{\partial \delta_c}
\end{equation}
where $R^{NG}(\fnl,M,z)=\frac{n^{NG}(\fnl,M,z)}{n^G(M,z)}$ denotes the ratio between the non-Gaussian mass function and the Gaussian one, that can be modeled in different ways \cite{matarrese_abundance_2000,loverde_effects_2008,maggiore_halo_2010,damico_improved_2011}.
The scale independent correction $\Delta b_I$ is due to the fact that the modulation induced by $f_{\rm NL}$ changes the mean number density of collapsed objects $\bar{n}$ compared to the Gaussian case.
 
The second correction regards the scale-dependent bias term mentioned above. It can be computed within the peak-background split formalism \cite{slosar_constraints_2008} by making use of Eq.~\eqref{eq:ds}, from which
\begin{equation}
 \Delta b(k,f_{\rm NL},M,z)=2 f_{\rm NL}\frac{d\phi_l}{d\delta_l}\frac{\partial \ln n^G}{\partial (1+2 f_{\rm NL}\phi_l)}=\frac{2f_{\rm NL}}{{\cal M}(k,z)}\frac{\partial \ln n^G}{\partial \ln \sigma_8^{\rm local}}
\label{eq:deltabk}
\end{equation}
where we have used the fact that locally, the effect of non-Gaussianity is a rescaling of the small scale matter fluctuations $\sigma_8 ^{\rm local}=(1+2f_{\rm NL}\phi_l)\sigma_8$ and thus  $d \sigma_8^{\rm local}=\sigma_8 d(1+2 f_{\rm NL}\phi_l)$. This is known as separate universes argument.

This correction becomes dominant at large scales, due to the fact that $T(k\rightarrow0) = 1$ and thus ${\cal M}(k,z)\propto k^2 D(z)$. Its amplitude is determined by the bias coefficient $b_{\phi}$ which, as we will show below, depends on the properties of the population of tracers considered. By comparing Eq.~\eqref{eq:dh_expansion} with \eqref{eq:deltabk} we can identify $b_{\phi}$ as
 \begin{equation}
 b_\phi=2\frac{\partial \ln n}{\partial \ln \sigma_8}.
 \label{eq:bphi}
\end{equation}
where we have dropped the G superscript and the \textit{local} label for brevity. As pointed out by \cite{slosar_constraints_2008}, we note that although $b_{\phi}$ drives the PNG-dependent contribution to the bias, it is expressed entirely in terms of Gaussian quantities, except for a cosmology with a suitably rescaled $\sigma_8$.

\subsection{$b_{\phi}$ and the universality relation}

As shown above, in order to provide predictions for $b_{\phi}$, it is sufficient to specify the halo mass function and measure its response to changes in $\sigma_8$. The standard result is obtained assuming a universal mass function \cite{slosar_constraints_2008,bond_excursion_1991,press_formation_1974,sheth_large_1999}
\begin{equation}\label{eq:hmf}
n(M,z) = \frac{\bar{\rho}}{M^2}\nu \mathfrak{f}(\nu) \frac{d \ln \nu}{d \ln M}
\end{equation}
where $\bar{\rho}$ is the background density, $\nu = \frac{\mathfrak{d}_c(z)}{\sigma(M)}$ is the ratio between the redshift-dependent collapse threshold $\mathfrak{d}_c(z)=\frac{\delta_c}{D(z)}$ and $\sigma(M)$, the amplitude of fluctuations in the linear density field, given by
\begin{equation}\label{eq:variance_pk}
S(M) \equiv \sigma^2 (M) = \frac{1}{2\pi^2}\int_0^{\infty} P_{\rm lin}(k)W^{2}_{M}(k)k^2 dk 
\end{equation}
where $P_{\rm lin}(k)$ is the power spectrum of the linear density field, which is  smoothed with a top-hat filter $W_M(k)$ on a scale enclosing the mass $M$.
In the Press-Schechter (PS) formalism,  the number density of collapsed objects $n(M,z)$ is stated in terms of a first-crossing problem: at each spatial point, the trajectory $\delta(M)$ of the smoothed linear matter density as a function of the smoothing scale $M$ undergoes a random walk; the upcrossing of the barrier $\mathfrak{d}_c(z)$ corresponds then to the collapse of matter into halos. The solution of this problem is given by the first-crossing probability distribution $\mathfrak{f}(\nu)$. A number of different functional forms of $\mathfrak{f}(\nu)$ have been proposed in the literature \cite{press_formation_1974,sheth_large_1999,sheth_ellipsoidal_1999,jenkins_mass_2000}, depending on the choice of the filter and the barrier. However, its exact form does not need to be specified to compute $b_{\phi}$, as long as it is \textit{universal}, i.e., only depends on mass and redshift via the variable $\nu$. For such a mass function, the prediction for $b_{\phi}$ (using the $u$ superscript to indicate the assumption of a universal mass function) is \cite{slosar_constraints_2008}
\begin{equation}
b^{u}_{\phi}=2 \delta_c (b_1-1)
\label{eq:univ_relation}
\end{equation}
and agrees with previous predictions derived independently \cite{dalal_imprints_2008,matarrese_effect_2008}.

As remarked in a number of recent works \cite{lazeyras_assembly_2022,barreira_impact_2020,barreira_galaxy_2020,barreira_can_2022} and already pointed out in the seminal papers in the field \cite{slosar_constraints_2008,seljak_measuring_2009,slosar_optimal_2009,reid_non-gaussian_2010}, this result suffers from two main problems.

First, the value of $\delta_c$ depends on the assumptions made about the formation (collapse) of halos. For example, assuming a spherical collapse, as considered in the PS formalism \cite{press_formation_1974}, $\delta_c = 1.686$. This assumption, however, might not be robust and motivated the development of non-spherical collapse models, characterized by a non-constant barrier \cite{sheth_ellipsoidal_1999,paranjape12,paranjape13,castorina16}. Furthermore,  as these predictions are  tested against N-body simulations, the actual definition of halos also depends on the  halo finder algorithm considered (as FoF, SO), which affects the halo number counts and thus their bias. Therefore, in order to include all these effects in a ``fudge factor" $q$, a modification $\delta_c \rightarrow q \delta_c$ is often considered\cite{grossi_large-scale_2009,desjacques_scale-dependent_2009,pillepich_halo_2009,Wagner:2011wx,biagetti_verifying_2017}.

Moreover, and crucially in view of comparisons with observations, the result is based on the assumption of having a fair sample of all the halos in a given mass range.  Equation \eqref{eq:univ_relation} does not hold, for example, for a sample of objects populating only recently merged halos \cite{slosar_constraints_2008}. Therefore, it has been argued that it should be rather modified as
\begin{equation}
    b_{\phi}^s=2\delta_c(b_1-p)
    \label{eq:bphip}
\end{equation}
where we explicitly indicate that $b_{\phi}$ is specific to the sample $s$, and the sample characteristics define the value of $p$, with $p=1$ for a fair sample of halos and $p=1.6$ for recent mergers \cite{slosar_constraints_2008,seljak_measuring_2009}.

\subsection{Halo assembly bias and its relation with $b_{\phi}$}

Dark matter halo tracers (e.g., galaxy) properties are well known to depend primarily on the host  halo mass, however there are secondary parameters modulating galaxy formation, such as environment and  halo formation history. If the selection of a sample of tracers depends crucially on the halo formation history (e.g., quasars, that are triggered by recent mergers), then Eq.~\eqref{eq:univ_relation} will not hold.

Following \cite{slosar_constraints_2008,reid_non-gaussian_2010} we explicitly recognize that in the extended Press-Schechter (ePS) approach \cite{lacey_merger_1993,lacey_merger_1994,bosch_universal_2001} it is possible to use the conditional mass function to make explicit the dependence of  $b_{\phi}$ on the halo formation time. This can be done by considering the number density of halos of mass $M_o$, observed at redshift $z_o$, that had accreted\footnote{Here we do not distinguish between  major/minor mergers or continuous accretion, but see e.g., \cite{AlizadehWandelt}.} a fraction $f$ of their final mass by formation time $z_f$. The $b_{\phi}$ associated to such halos is given by
\begin{equation}
 b_\phi (M_o,z_o,z_f,f)=2\frac{\partial \ln n(M_o,z_o)}{\partial \ln \sigma_8}+2\frac{\partial \ln P_{z_f}(fM_o,z_f|M_o,z_o)}{\partial \ln \sigma_8}=b_{\phi}^u+\Delta b_{\phi},
 \label{eq:eps}
\end{equation}
which follows from Eq.~\eqref{eq:bphi}. $P_{z_f}$ denotes conditional mass function: the probability that a halo with mass $M_o$ at $z_o$ had a mass $fM_o$ in a redshift interval $dz_f$ around a higher redshift $z_f$.

The first term, under the assumption of a universal mass function, reduces to Eq.~\eqref{eq:univ_relation}, hence if the generic parameterization of $b_{\phi}$ in terms of $p$ is considered, as in Eq.~\eqref{eq:bphip}, we can identify the second term with $\Delta b_{\phi}=2\delta_c (1-p)$. Thus the conditional mass function, i.e., the merger history of the objects selected, can be used to put a prior on the value of $p$, or more generally on $b_{\phi}$. For a fair sample of halos of mass $M$,  $\Delta b_{\phi}=0$ and $p=1$, otherwise it is a potentially very important correction \cite{reid_non-gaussian_2010,lazeyras_assembly_2022,barreira_impact_2020,barreira_galaxy_2020,barreira_can_2022}.

As mentioned above, in the PS formalism the halo mass function is expressed in terms of the distribution $\mathfrak{f}(\nu)$ of first crossings of a barrier $\mathfrak{d}_c(z)$. Following \cite{lacey_merger_1993,lacey_merger_1994,bosch_universal_2001}, the conditional mass function can be computed in the same way, in the context of a diffusion problem between two barriers $\mathfrak{d}_c(z_o)$ and $\mathfrak{d}_c(z_f)$. Since for random walks the crossing probability does not depend on the path, the solution of this problem has the same form as $\mathfrak{f}(\nu)$, with a transformation of the origin of coordinates. In the specific case of the PS solution \cite{press_formation_1974}, expressed in terms of $S_f=S(fM_o)$ and $S_o=S(M_o)$ as defined by Eq.~\eqref{eq:variance_pk}, we have 
\begin{equation}\label{eq:press_schechter_solution}
    \frac{d \mathfrak{f}}{d S_o}(S_f,z_f|S_o,z_o) = \frac{1}{\sqrt{2\pi}}\frac{\mathfrak{d}_c(z_f)-\mathfrak{d}_c(z_o)}{(S_f-S_o)^{3/2}}e^{-\frac{[\mathfrak{d}_c(z_f)-\mathfrak{d}c(z_o)]^2}{2(S_f-S_o)}}
\end{equation}
which gives the conditional probability for objects of a given mass $M_o$. Then, we use Eq.~\eqref{eq:hmf} to convert mass fractions to number of halos and define
\begin{equation}\label{eq:omega_f}
    \omega_f\equiv\frac{\mathfrak{d}_c(z_f)-\mathfrak{d}(z_o)}{\sqrt{\sigma^2(fM_o)-\sigma^2(M_o)}},
\end{equation}
which is the variable that we use to parameterize assembly bias and is proportional to $z_f-z_o$ at fixed $M_o$, as described in detail in Sec.~\ref{ssec:omega_f}. After some calculations, we obtain the conditional probability distribution
\begin{equation}\label{eq:P_omega_f}
   P_{\omega_f}(\omega_f)=\frac{1}{\sqrt{2\pi}}\int^1_0 \frac{M_o}{M(\tilde{S})}\left(\frac{\omega_f^2}{S^{5/2}}-\frac{1}{S^{3/2}}\right)e^{-\frac{\omega_f^2}{2S}}dS,
\end{equation}
with $\tilde{S}=(S_f-S_o)S+S_o$.

Implementing this change of variable in Eq.~\eqref{eq:eps}, the prediction for $\Delta b_{\phi}$ can be also expressed in terms of a derivative with respect to $\omega_f$:
\begin{equation}
\Delta b_{\phi}=2\frac{\partial \ln P_{z_f}(fM_o,z_f|M_o,z_o)}{\partial \ln \sigma_8} =-2\left(1 +\frac{d\ln P_{\omega_f}}{d\ln {\omega_f}}\right).
\label{eq:delta_bphi}
\end{equation}
Consequently, if one considers the parameterization of $b_{\phi}$ through the variable $p$, as in Eq.~\eqref{eq:bphip}, then the expression for $p$ becomes
\begin{equation}
 p=1+\frac{1}{\delta_c}\left(1+\frac{d\ln P_{\omega_f}}{d\ln {\omega_f}} \right)\,.
 \label{eq:peff}
\end{equation}
Therefore, the correction to the universality relation, computed by including the effect of assembly bias, depends on redshifts ($z_f$ and $z_o$) and mass ($M_o$ and $fM_o$) only through the single variable $\omega_f$.

The approach considered here and the analytical expression of $P_{\omega_f}$ in Eq.~\eqref{eq:P_omega_f} provide a remarkably good description at high halo masses ($M_o\gtrsim 5\times 10^{12} M_{\odot}$), as already tested in previous works \cite{reid_non-gaussian_2010,bosch_universal_2001}. This result is further corroborated by  comparisons with different N-body simulations below. However, for lower halo masses, the analytical prediction breaks down, due to $P_{\omega_f}$ taking a dependence on $M_o$ and $z_o$.

We highlight here that the form of Equations \eqref{eq:delta_bphi} and \eqref{eq:peff} does not depend on the assumptions made about the analytical form of $P_{\omega_f}$. In fact, they could directly follow from Eq.~\eqref{eq:eps} by adopting the change of variable defined by Eq.~\eqref{eq:omega_f}. Therefore, a prediction for $p$, or equivalently for $\Delta b_{\phi}$, can be obtained from $P_{\omega_f}$ even if its expression is calibrated on dark matter simulations and not derived analytically. Building upon this argument, we provide below a simple modification of the ePS formula (which we refer to as e$^2$PS) for $P_{\omega_f}$, which fits  also the simulated low mass halos well.

\section{N-body simulations and comparison with ePS-derived  expressions}
\label{sec:ePStest}
In this work, we consider different sets of simulations with a double aim: on one hand, we make use of N-body simulations to evaluate the accuracy of the theoretical predictions described above; on the other hand, we study through hydrodynamical simulations how galaxy properties might be used as a proxy for halo assembly bias. To this end, we consider both the halo catalogs and halo merger trees available in these sets, which allow us to study the $b_{\phi}$ dependence on the halo formation history.

One set is the Quijote-PNG \cite{coulton_quijote_2022} suite of gravity-only simulations, which includes primordial non-Gaussianity of the local type in the initial conditions.
This suite of N-body simulations consists in 1000 different realizations, each of them having a different value of local $f_{\rm NL} \in [-300,300]$ and different seed of the initial conditions. The cosmological parameters are fixed to match those used in the fiducial Gaussian suite Quijote \cite{villaescusa-navarro_quijote_2020}, which we also use in our work. The size of the simulation box is $L=1 \,h^{-1}$Gpc and the particle mass $m_p = 6.6 \times 10^{11} h^{-1} M_\odot$.

Another set is the IllustrisTNG \cite{nelson_illustristng_2021} suite of high-resolution magneto-hydrodynamical simulations with Gaussian initial conditions. These simulations use a model for galaxy formation and evolution, calibrated to reproduce a number of galaxy properties such as the star formation rate, photometry and metallicities. Specifically, we focus on the TNG300 simulation, which uses a box with size $L=205\,h^{-1}$Mpc and has a mass resolution of $m_p = 7 \times 10^{7} h^{-1} M_\odot$. In this section we make use  only of the dark matter aspect of the IllustrisTNG, the baryonic (galaxy properties) part is considered in Section \ref{sec:proxy}.  

Table \ref{tab:simulation_params} summarizes the main properties of the simulations, including their cosmological parameters.
The mass resolution introduces a mass limit under which the halos are not well resolved and their formation redshift is not well defined. This is due to the fact that for small $M_o$, there may not exist any progenitor with mass $fM_o$ (from which $z_f$ is defined) across the merger tree. Therefore, in the following we will consider only halos above a certain mass threshold: $M_o>10^{14} h^{-1} M_\odot$ for Quijote-PNG and $M_o>5\times 10^{11} h^{-1} M_\odot$ for IllustrisTNG. These thresholds guarantee that $z_f$ is well defined for all the halos at $z_o=0$ and are sufficient for $z_o>0$ halo samples. Due to the different mass resolution, the two simulations can be used  in a complementary way to test the ePS predictions on different mass ranges.

\begin{table}
    \centering
    \begin{tabular}{c c c c c c c c c c}
 & $\Omega_m$ & $\Omega_\Lambda$  & $ \Omega_b $ & $ \sigma_8 $  & $h$ & $n_s$ & L & $m_p$ \\ 
  & & & &  & & &  (Mpc/h) & (M$_\odot$/h)\\
    \hline \hline 
 Quijote-PNG & 0.3175 & 0.6825 & 0.049 & 0.834 & 0.6711 & 0.9624 &  1000 & $6.6\times10^{11}$\\
 IllustrisTNG & 0.3089 & 0.6911 & 0.049 & 0.816 & 0.6774 & 0.9667 &  205 & $7\times10^{7}$ \\
   \end{tabular}
    \caption{Summary of the cosmological parameters and properties of the simulations used in the analysis.\label{tab:simulation_params}}
\end{table}

In order to study the predictions on the halo merger history, we make use of Quijote-PNG (IllustrisTNG) halo catalogs constructed with the Rockstar (FoF) halo finder and merger trees generated with the ConsistentTrees (Sublink) algorithm. For each halo identified in the catalog at $z_o$, its mass $M(z)$ is tracked back along the main (most massive) progenitor branch. The mass history is then reconstructed by interpolating $M(z)$ across the available snapshots and the formation redshift $z_f$ is defined as the redshift at which the halo mass is $M(z_f)=fM_o= M_o/2$, with $M_o$ being the  mass at $z_o$. Hence each halo identified at $z_o$ is associated a mass $M_o$ and a formation redshift $z_f$. Then, the variable $\omega_f$ is computed for each halo by using Eq.~\eqref{eq:omega_f}.

Our choice of $f=0.5$ follows the original prescription of the ePS  presented in Ref.~\cite{lacey_merger_1993}. Although different values, as for example $f=0.75$, have been chosen in the literature to describe cluster abundances \cite{viana96}, Ref.~\cite{reid_non-gaussian_2010} showed that, in the context of the non-Gaussian halo assembly bias, the two choices give consistent results. More generally, the description of assembly bias may be improved by considering other, integrated,  quantities beyond the formation redshift $z_f$, which is subject to the variability of the halo mass accretion history, and thus might be noisy.  In this particular context,  especially  where we consider large halo samples, the noise gets effectively averaged out.  The  parameterization chosen here benefits from having a transparent relation with the analytical prediction given by the ePS. Hence the adoption of less noisy parameterizations is left to future work.

\subsection{Describing halo formation history through $\omega_f$ and$P_{\omega_f}$}\label{ssec:omega_f}
In the context of the ePS formalism, the variable $\omega_f$ was introduced in order to express the conditional mass function in terms of a single variable, as $P_{\omega_f}(\omega_f)$. Strictly speaking, this is only valid under the approximation of an Einstein de Sitter cosmology with a power law linear matter power spectrum, as the one adopted in \cite{lacey_merger_1993,lacey_merger_1994}.

In fact, we see from Eq.~\eqref{eq:P_omega_f} that the ePS prediction also depends on the ratio $\frac{M_o}{M(\tilde{S})}$ which introduces a mass dependence, although it is very mild. Only by knowing the relation $S(M)$ analytically, this ratio can be integrated out and $P_{\omega_f}$  depends on $\omega_f$ only. Specifically, Ref. \cite{lacey_merger_1993,lacey_merger_1994} consider an EdS universe, where $D(z)=(1+z)^{-1}$ and assume a power-law linear matter power spectrum $P_{lin}(k) \propto k^n$. In this case, the variance can be analytically computed from Eq.~\eqref{eq:variance_pk} and written as $S(M)=\delta_c^2\left(\frac{M_{*}}{M}\right)^{\frac{n+3}{3}}$, where $M_{*}$ is a characteristic mass such that $S(M_{*})=\delta_c^2$.

\begin{figure}[H]
\begin{center}
	\includegraphics[width=0.7\linewidth]{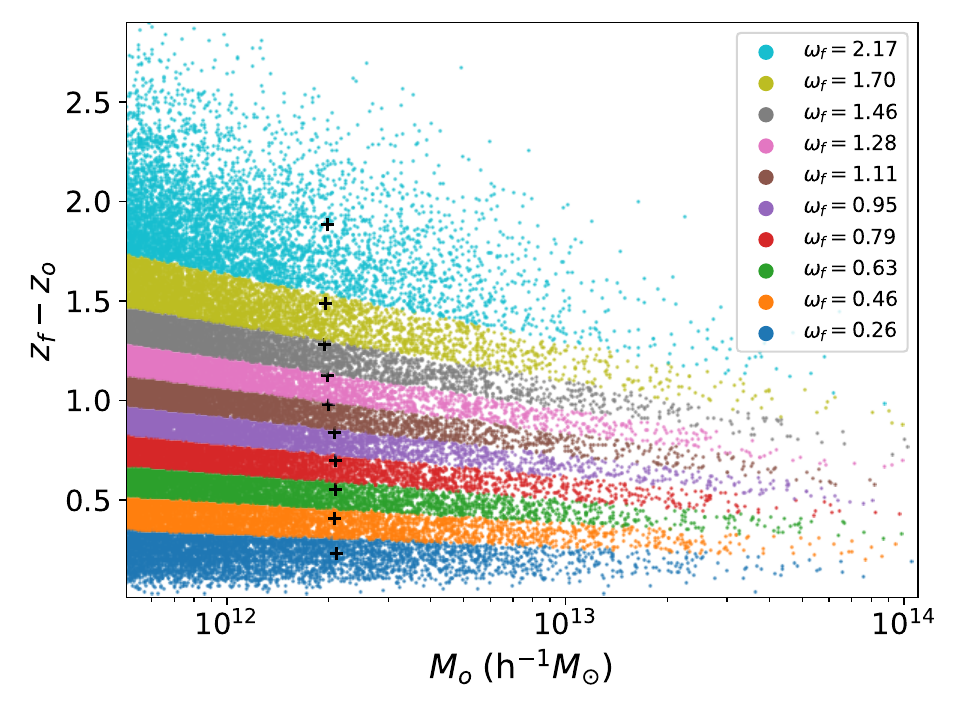}
\end{center}
 \caption{$z_f-z_o$ as a function of halo mass for the full sample of IllustrisTNG halos at $z_o$=1. The halos are sorted by their $\omega_f$ and divided in subsamples. Each subsample, represented by a different colour, contains 10\% of the full halo sample. The black crosses indicate the mean mass and mean $z_f-z_o$ of the subsamples. The mean value of $\omega_f$ in each subsample is reported in the legend.}
 \label{fig:wf_subs}
\end{figure}

In this approximation, we can better understand the relation between $\omega_f$ and $M_o,z_o,z_f$. By substituting these approximations in Eq.~\eqref{eq:omega_f}, we obtain
\begin{equation}
\label{eq:omega_f_approx}
    \omega_f = \left(f^{-\frac{n+3}{3}}-1\right)^{-1/2}\left(\frac{M_o}{M_{*}}\right)^{\frac{n+3}{6}}(z_f-z_o)
\end{equation}
 \begin{figure}[H]
\begin{center}
	\includegraphics[width=0.7\linewidth]{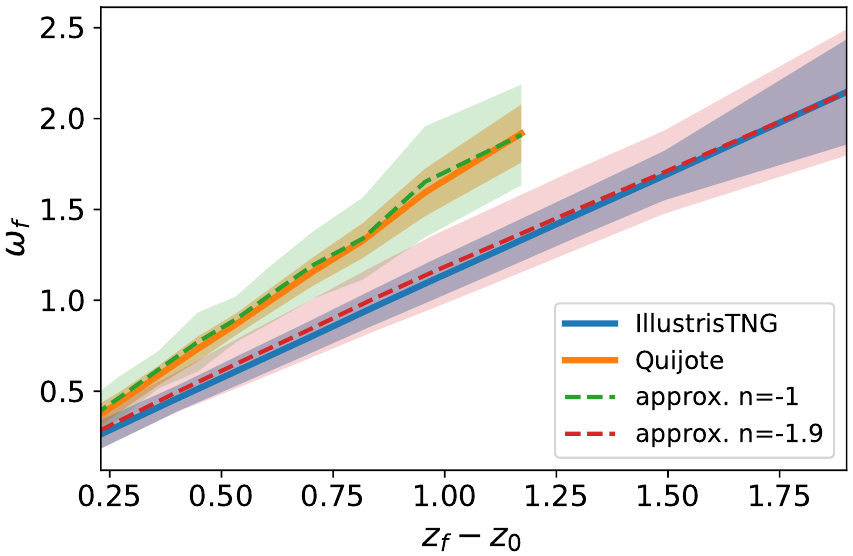}
\end{center}
 \caption{Relation between $\omega_f$ and $z_f-z_o$ for IllustrisTNG ($\bar{M}=2\times 10^{12} M_{\odot}{\rm h^{-1}}$) and Quijote ($\bar{M}=2\times 10^{14} M_{\odot}{\rm h^{-1}}$) halos at $z_o=1$. The lines and the bands represent the mean and the standard deviation $\omega_f$ computed in $z_f-z_o$ bins of full halo sample. We show both the approximated $\omega_f$, computed through Eq.~\eqref{eq:omega_f_approx}, and the actual one, computed using Eq.~\eqref{eq:omega_f}. The values of $n$ are obtained by fitting the approximation against the actual values of $\omega_f$ in the two samples.}
 \label{fig:wf_zf}
\end{figure}
Therefore at fixed mass $M_o$, $\omega_f$ is a linear rescaling of $(z_f-z_o)$, with a slope dependent on the power spectrum exponent $n$.
The power-law approximation is valid in a sufficiently small range of $k$, which translates into a range of halo masses through the top-hat filter smoothing.
Therefore, the power-law index $n$ will vary according to the mean mass $\bar{M_o}$ of the halo sample considered. In Fig.~\ref{fig:wf_zf} we compare this approximation with the original expression in Eq.~\eqref{eq:omega_f} for both IllustrisTNG ($\bar{M_o}=2\times 10^{12} M_{\odot}{\rm h^{-1}}$) and (Gaussian) Quijote ($\bar{M_o}=2\times 10^{14} M_{\odot}{\rm h^{-1}}$) halos at $z_o=1$. For these two different mass ranges, the approximation works well for a power-law power spectrum with $n=-1.9$ and $n=-1$, respectively. The scatter around the mean linear relation is due to the fact that the halo mass $M_o$ varies inside the samples considered.
The maximum $z_f-z_o$ in  the Quijote sample is lower than the IllustrisTNG one, reflecting the fact that heavier halos tend to form later than lighter ones, as can be noticed by looking at Fig.~\ref{fig:wf_subs}.
In what follows, unless specified, we will always use the Eq.~\ref{eq:omega_f} to compute $\omega_f$, but this linear relation between $\omega_f$ and $z_f-z_o$ indicates that, in small mass bins, selecting a certain fraction of the oldest (youngest) halos is equivalent to selecting the same fraction of halos with largest (smallest) $\omega_f$, as assumed by Ref. \cite{reid_non-gaussian_2010}.
However, $z_f$- and $\omega_f$- subsamples selected from the full sample of halos do not exactly match. 
In Fig.~\ref{fig:wf_subs} we illustrate how quantiles of halos selected by their $\omega_f$ value, each one containing 10\% of the full halo sample, project onto the $(M_o,z_f-z_o)$ plane. Generally, the same $\omega_f$ quantile contains both low mass old halos and higher mass younger halos. This ``tradeoff" can be understood by looking at the approximation \eqref{eq:omega_f_approx} for $\omega_f$ fixed. Incidentally, due to this tradeoff, the different $\omega_f$ quantiles share roughly the same mean halo mass (differing at most by 3\%).

\begin{figure}[H]
    \begin{minipage}{.65\textwidth}
	\includegraphics[width=1\linewidth]{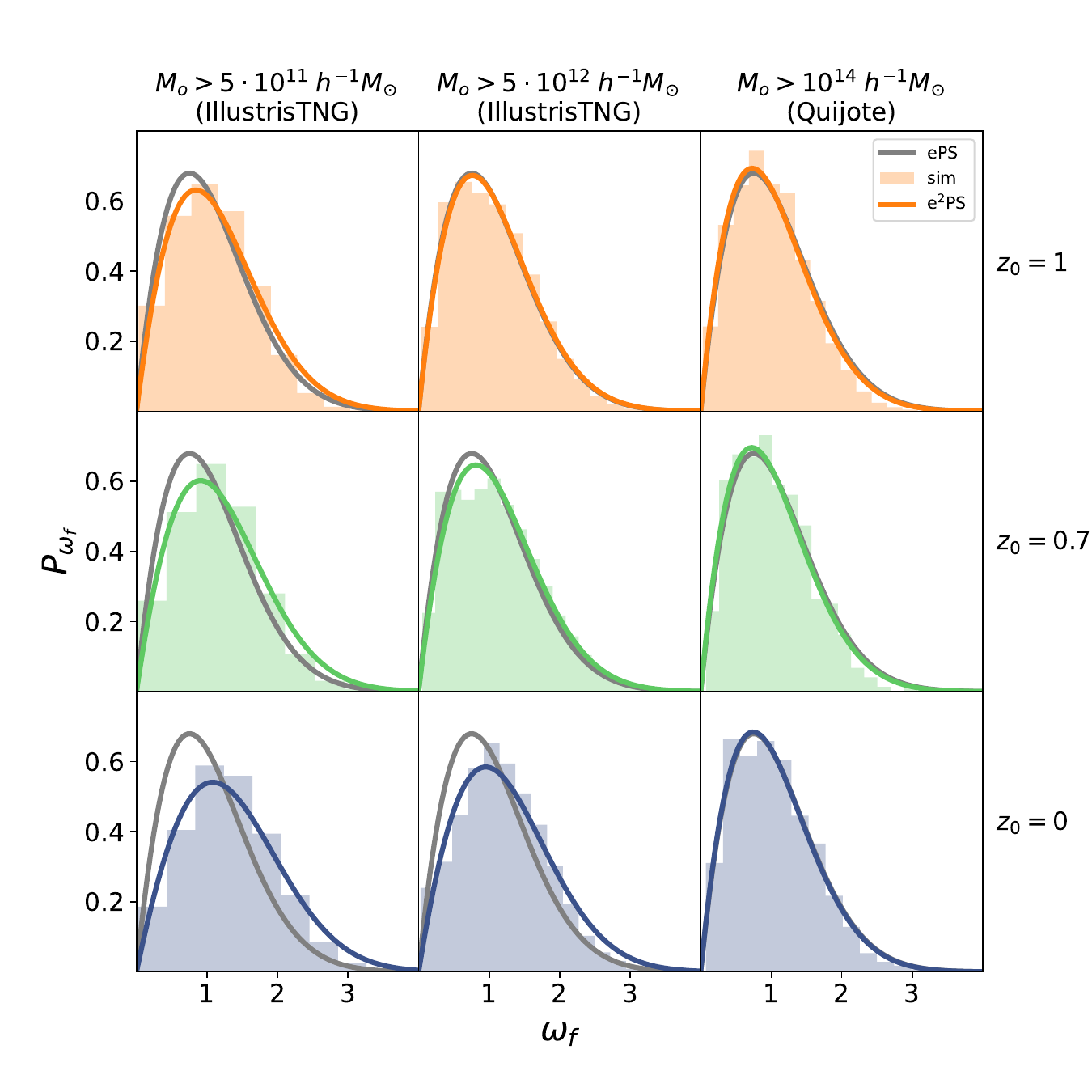}
    \end{minipage}
    \begin{minipage}{.33\textwidth}
    \vspace{0.3cm}
	\includegraphics[width=1\linewidth]{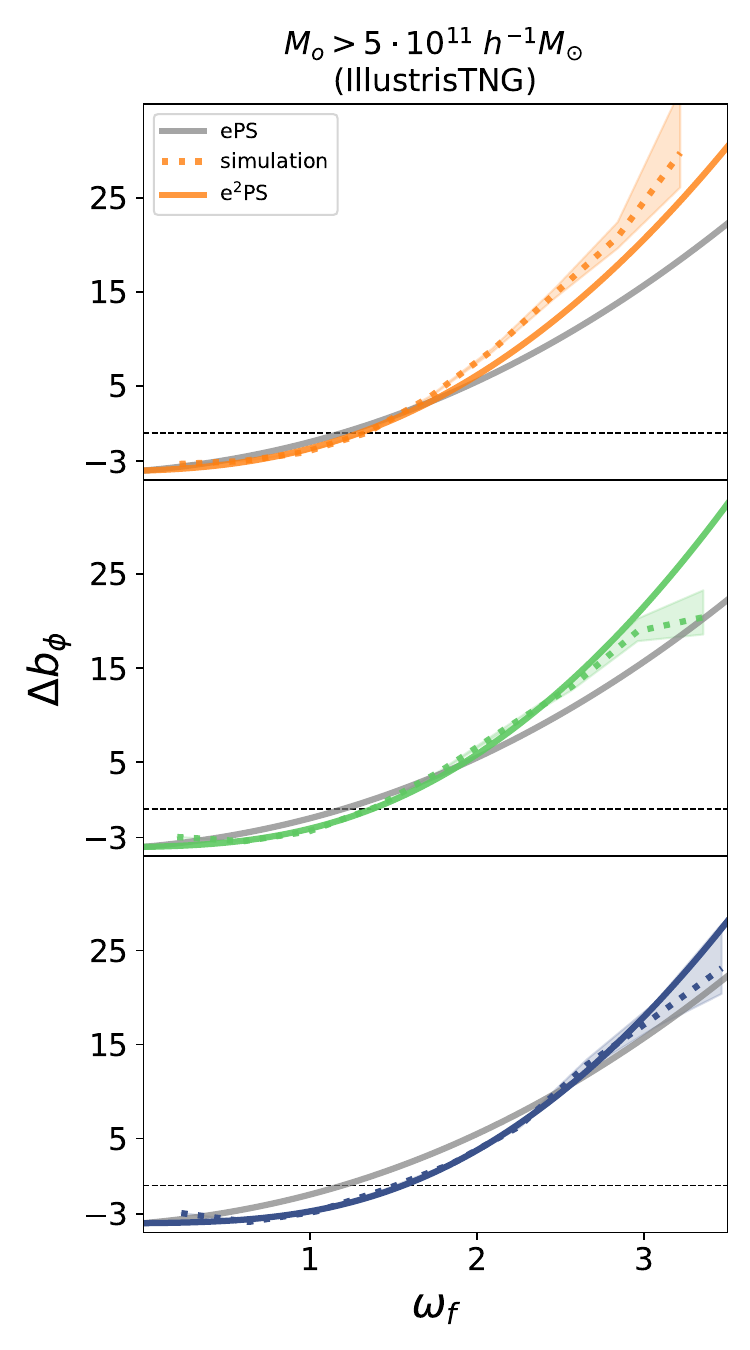} 
    \end{minipage}

	\caption{Comparison between results from N-body simulations and the predictions from ePS and e$^2$PS. The grid on the left shows the results for $P_{\omega_f}$. The mass $M_o$ of the halos considered increases from left to right, while $z_o$ increases from bottom to top. The first two columns show the distributions of IllustrisTNG halos, while the third one for Quijote. In each panel, the histograms represent the results from simulations, the grey curve corresponds to the ePS prediction, while the coloured one is the best-fit e$^2$PS curve. On the right, we show the $\Delta b_{\phi}$ computed from Eq.~\eqref{eq:delta_bphi}. The solid gray (coloured) line correspond to the ePS (e$^2$PS) model, the dotted lines correspond to the simulations. The coloured band around the simulation curve quantifies the error propagated from the bin count errors in the corresponding histogram on the grid. For brevity, we only report the results related to the first column of the grid. The dashed black horizontal line corresponds to the universality relation prediction, $\Delta b_{\phi}=0$. }
	\label{fig:eps_test}
\end{figure}

For the approximation described above, and for $f=0.5$, we can solve the integral in Eq.~\eqref{eq:P_omega_f} analytically and obtain the following form for $P_{\omega_f}$:
\begin{equation}\label{eq:pwf_eps}
P_{\omega_f}=2\omega_f {\rm erfc}\left(\frac{\omega_f}{\sqrt{2}}\right)
\end{equation}
where ${\rm erfc}(\omega_f)=1-{\rm erf}(\omega_f)$ is the complementary error function. The function in Eq.~\eqref{eq:pwf_eps} only depends on $\omega_f$ and as such it should be universal, i.e., valid regardless the mass and the redshift of the halos considered.
Although this is true at large halo masses \cite{reid_non-gaussian_2010}, a comparison with the $P_{\omega_f}$ constructed from the simulations considered in the present work shows a residual mass and redshift dependence. In particular, as we show in Fig.~\ref{fig:eps_test}, as the mean mass and the redshift of the halo sample decrease, the ePS prediction does no longer accurately describe the $P_{\omega_f}$ obtained from simulations.

However, the shape of the distribution is preserved, although broadened and with a mean shifted towards larger $\omega_f$ values (i.e., earlier formation redshift). In order to account for this dependence, we propose a simple modification of Eq.~\eqref{eq:pwf_eps}, shifting the argument of the ${\rm erfc}$ by a constant $\omega_0$:
\begin{equation}\label{eq:pwf_e2ps}
    P_{\omega_f}=P_{\omega_f}(\omega_f|\omega_0)=N(\omega_0)\omega_f{\rm erfc}\left(\frac{\omega_f-\omega_0}{\sqrt{2}}\right),
\end{equation}
where we fit for the value of $\omega_0$, dependent on the mass and redshift of the sample of halos considered. The factor $N(\omega_0)$ ensures that the probability distribution is correctly normalized (and is such that $N(\omega_0=0)=2$). The functional form of Eq.~\eqref{eq:pwf_e2ps}, which we will refer to as e$^2$PS for the rest of the paper, can be considered as a phenomenological model, or a fitting formula which includes the ePS prediction as a particular case with $\omega_0=0$.
The e$^2$PS formula shows a remarkably good fit with simulations across different redshifts and mass ranges, as shown in Fig.~\ref{fig:eps_test}.

The best-fit value of $\omega_0$ approaches $\omega_0=0$ as the mass and the redshift increase and correspondingly, the e$^2$PS starts to coincide with the ePS, as illustrated in Fig.~\ref{fig:eps_test}.

In the right panel, we report the prediction for $\Delta b_{\phi}$ as computed from Eq.~\eqref{eq:delta_bphi} by taking the derivative of the conditional mass function. We show both the ePS prediction and the one obtained from simulations, by taking the numerical derivative of the histograms reported on the left. Moreover, we also report the prediction for $\Delta b_{\phi}$ as derived from the best-fit e$^2$PS curve, calibrated on the corresponding $P_{\omega_f}$. The prediction for large values of $\Delta b_{\phi}$ is less precise, as it is related to the tail of the distribution $P_{\omega_f}$, where the noise increases due to small number statistics.

At redshifts snapshots $z=0$ and $z=0.7$ there is no evidence for any significant bias induced in $b_{\phi}$ by the e$^2$PS predictions keeping in mind that  extremely large $\omega_f$ and $\Delta b_{\phi}$ values  apply only to extremely rare objects with large statistical errors. At $z=1$ there may be a hint of a possible  mis-match  for $\Delta b_{\phi}>5$, $\omega_f>2$, but statistics are poor; we return to this point in Fig.~\ref{fig:b_phi}. 

The mass and redshift dependence of $P_{\omega_f}$ is a result of the breakdown of the universality of the mass function, i.e., the Press-Schechter solution \eqref{eq:press_schechter_solution}. This effect has already been discussed in the literature \cite{tinker_toward_2008,pillepich_halo_2009} and modifications of the assumptions made in the PS approach might explain this discrepancy. This is beyond the scope of this work, but some insights on the theoretical modeling can be gleaned by looking at the form of the e$^2$PS formula \eqref{eq:pwf_e2ps}. In particular, the presence of a correction $\omega_0=\omega_0(M,z)$ suggests the modification $\mathfrak{d}_c(z)\rightarrow\mathfrak{d}_c(M,z)$ in Eq.~\eqref{eq:omega_f}. Such a modification would require the determination of the upcrossing statistics for what is known in the literature as a \textit{moving barrier}\cite{sheth_ellipsoidal_1999,maggiore_halo_2010,paranjape12,paranjape13,castorina16,lazeyras17}.

\subsection{Accuracy of $P_{\omega_f}$ modeling: Results on $\Delta b_{\phi}$ from non-Gaussian simulations}
In the previous section we have shown how the variable $\omega_f$ and its distribution $P_{\omega_f}$ can be used to study halo assembly bias, reporting results from Gaussian simulations. Once $P_{\omega_f}$ is known, the deviation from the universality relation, $\Delta b_{\phi}$, can be obtained from Eq.~\eqref{eq:delta_bphi}. Beyond being measured from simulations, these quantities have been tested against the ePS prediction and a fitting function has been provided when this prediction breaks down.
In this section we validate these results on the Quijote-PNG non-Gaussian simulations, fitting $\Delta b_{\phi}$ by measuring the scale dependent bias induced by local PNG.

Of the Quijote-PNG suite we consider 500 simulations with $|\fnl|>150$. This selection serves to amplify the signal and obtain better constraints on $b_{\phi}$.

For each simulation, we study the two snapshots at $z_o=0, 1$. After having identified each halo in the snapshot, its $z_f$ is computed by interpolating across the merger tree and the pair of variables $(M_o,z_f)$, where $M_o$ is the halo mass,  are transformed into the variable $\omega_f$ by using Eq.~\eqref{eq:omega_f}.

Our aim is to divide the full halo sample in each snapshot in 10  $\omega_f$ bins  so that each bin has the same number of halos, as those in Fig.~\ref{fig:wf_subs}. As it is evident from Fig.~\ref{fig:wf_subs} even though these subsamples have similar mean masses, it is not guaranteed they would have matching mass functions. As we can see from Eq.~\eqref{eq:eps}, this is required to isolate the universality relation contribution to $b_{\phi}$ and fit $\Delta b_{\phi}$, as also discussed in \cite{reid_non-gaussian_2010}. Therefore, following \cite{reid_non-gaussian_2010}, we proceed as follows.

We first divide the full sample in 30 small mass bins, each one containing the same number of halos (and having larger width for larger masses, accordingly to the mass function). Each mass bin is then divided into 10 $\omega_f$ quantiles. Finally, halos belonging to the same $\omega_f$ quantile in different mass bins are stacked together. By construction, we end up having 10 samples with mass functions matching that of the full sample.

From these 10 samples, we construct 10\%, 20\%, ..., 90\% quantiles of the \textit{oldest} (large $\omega_f$) and \textit{youngest} (small $\omega_f$) halos. We test the predictions discussed above by analyzing the clustering properties of these quantiles, following the same approach as \cite{reid_non-gaussian_2010}. For each quantile, we construct the dark matter density field $\dm$ from the dark matter simulations snapshots and the halo number overdensity $\dh$ from the corresponding halo catalog.

The scale-dependent bias affects the largest scales, where the number of power spectrum modes (when averaged over $k$ bins) is relatively low. Therefore, rather than computing a power spectrum, we choose to fit Eq.~\eqref{eq:dh_expansion} mode by mode, up to $|{\bf k}|<0.03$ Mpc $h^{-1}$ and assume to have Poisson shot noise, $\left<\epsilon({\bf k}) \epsilon^{\star}({\bf k})\right> = \bar{n}^{-1}$, with $\bar{n}$ mean number density of the sample.

The variance about the model is 
\begin{equation}
\left<\left(\delta_h^{\star} \delta_m -  \left(b_1 + b_{\phi} \frac{\fnl}{{\cal M}(k,z_o)}\right)\delta_m^{\star} \delta_m\right)^2\right> = \frac{P_{mm}(k)}{\bar{n}},
\label{eq:dh_var}
\end{equation}
where the average is computed over the Poisson noise, assumed to be uncorrelated with $\dm$. 
In order to perform the fit, we fit for the values of $b_1$ and $b_{\phi}$ that minimize the $\chi^2$, computed as follows:
\begin{equation}
\label{chi2eq}
\chi^2 = \displaystyle\sum_{\vk}\frac{\left(Re[\delta_h^{\star} \delta_m] -  \left(b_1 + b_{\phi} \frac{\fnl}{{\cal M}(k,z_o)}\right)\delta_m^{\star} \delta_m\right)^2}{P_{mm}(k)/(2\bar{n})} 
\end{equation}
We only consider the real component of $\delta_h^{\star} \delta_m$, hence the noise is a factor of 2 smaller than in Eq.~\eqref{eq:dh_var}.

For each simulation with different $\fnl$ and for each $z_o$, the fit procedure is applied to the full sample and the 9 cumulative subsamples of both the oldest and youngest halos. Then, $\Delta b_{\phi}$ is estimated from the difference between the $b_{\phi}$ fitted from the full sample and the one fitted from the subsample:
\begin{equation}
    \Delta b_{\phi} = b_{\phi}^{full}-b_{\phi}^{sub}.
\end{equation}
Finally, we take the average of $\Delta b_{\phi}$ over the different simulations, at same redshift $z_o$.

\begin{figure}[b]
\includegraphics[width=1\linewidth]{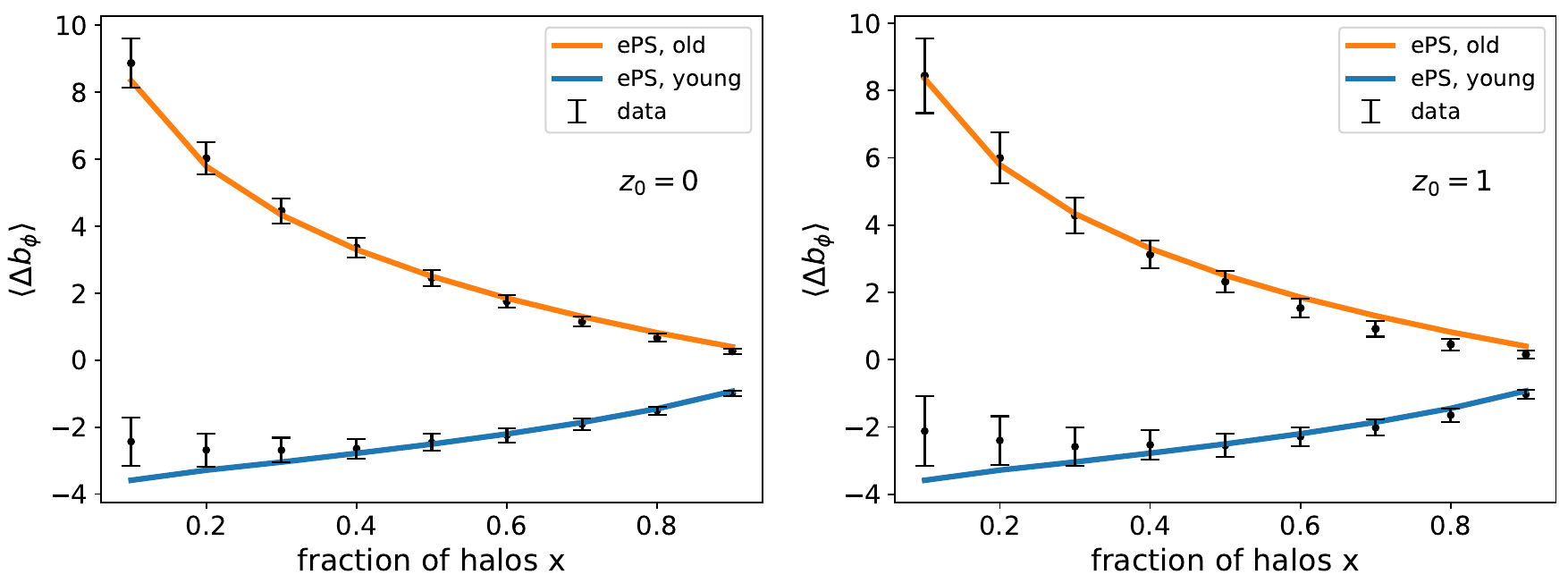}
\caption{Average $\Delta b_{\phi}$ measured from the cumulative subsamples of the youngest and oldest halos. ePS predictions for both fractions are also shown.  On the left, we show results for the population of halos at $z_o=0$. On the right, for $z_o=1$. The average is computed over 500 Quijote-PNG simulations with different $\fnl$ (and the errors denote the error on the mean). The halo mass threshold chosen to avoid unresolved halos is $M_o>10^{14}$ M$_\odot/h$.\label{fig:b_phi}}
\end{figure}

In Fig.~\ref{fig:b_phi} we report the results obtained at $z_o=0$ and $z_o=1$ for $\left<\Delta b_{\phi}\right>$, averaged over the 500 realizations with different $\fnl$. Each of the quantile we consider in the analysis extends across a range of $\omega_f$ values, $[\omega_f^{min},\omega_f^{max}]$. The theoretical prediction for the $\Delta b_{\phi}$ of each quantile is then found by averaging Eq.~\eqref{eq:delta_bphi}, 
\begin{equation}
\left<\Delta b_{\phi}\right>=-2\left<1 +\frac{d\ln P_{\omega_f}}{d\ln {\omega_f}}\right>,
\end{equation}
where the average is computed over the range $[\omega_f^{min},\omega_f^{max}]$, using the probability distribution $P_{\omega_f}$. In this case we use the ePS prediction for $P_{\omega_f}$, which, as discussed in the previous section, which is shown to be accurate for the range of halo masses of Quijote-PNG. 
The main result is that old halos have a larger $b_{\phi}$, while young ones have a lower $b_{\phi}$ with respect to that of the full sample. The effect is asymmetric between the two, as $\Delta b_{\phi}$ increases rapidly at large $\omega_f$, while gradually at small $\omega_f$, as illustrated in Fig.~\ref{fig:eps_test}. This asymmetry is in accordance with similar results which parameterized assembly bias through the halo concentration instead of the halo formation redshift \cite{lazeyras_assembly_2022,barreira_towards_2023,sullivan_learning_2023}.

If we consider the parameterization in terms of $p$, the 10\% of the oldest halos have a $\Delta b_{\phi}$ which corresponds to $p\simeq-1.4$, while the 10\% of youngest ones, $p\simeq1.6$, as already found in \cite{slosar_constraints_2008} for recent mergers. 

The ePS prediction shows a remarkable accuracy across the two different redshifts $z_o=0,1$, especially for the oldest halos;   there are some small discrepancies for small $\lesssim 0.2$ fractions of the youngest halos but the agreement is good for fractions $\gtrsim 0.3$. The errorbars are larger for $z_o=1$ as a result of the increased shot noise contribution with respect to $z_o=0$.
These results are consistent with those of \cite{reid_non-gaussian_2010}, where they used different non-Gaussian simulations and range of masses.

In summary, as Fig.~\ref{fig:b_phi} shows, there is no evidence of  significant systematic effect on $\Delta b_{\phi}$ introduced by the e$^2$PS modeling  above the statistical errors corresponding to a volume of 500 (Gpc/h)$^3$.

\section{Galaxy proxy for halo assembly bias: results from IllustrisTNG simulations}
\label{sec:proxy}
So far we have presented how to describe the impact of assembly bias on the local PNG bias parameter $b_{\phi}$, focusing on simulated halo catalogs and merger trees.
By knowing $(M_o,z_f)$  of a sample of halos, and computing $\omega_f$, we have shown how to obtain a prediction for $b_{\phi}$. However, halo properties such as $M_o,z_f$ are not  directly observable and their inference is tied to  the galaxy-halo connection, which description includes modeling halo occupation distribution (HOD), halo mass accretion history, as well as galaxy formation (see Ref. \cite{Wechsler_2018} for a detailed review).
Quantities which can actually be observed in LSS surveys are only (some) galaxy properties. Therefore, our aim is to find a galaxy-related quantity which can be connected to the host halo formation redshift $z_f$, or even better, to the $\omega_f$ of the corresponding host halo. In other terms, the goal is to find an observable proxy of halo assembly bias.

In the literature, different galaxy proxies have been used, most of which are also proxies for the age of the galaxy, which is expected to be statistically correlated to the age of the halo. Examples of properties used as proxies are stellar mass $M_{*}$\cite{Croton_2007}, specific star formation rate (sSFR)\cite{Watson_2014,Tinker_2018} and the galaxy color $g-r$\cite{Hearin_2013}. However, some of these properties are to some extent model-dependent: for example $M_{*}$ and sSFR of galaxies are indirect measurements inferred from spectroscopic and photometric data \cite{Hopkins_2003,Mitchell_2013}.

Our approach here is to remain as close  to observations as possible, focusing on assembly bias proxies constructed from (combination of) photometric bands beyond the optical ones $g$, $r$. While the proxy could certainly  be improved by adding extra information e.g., a detailed spectroscopic analysis, photometry is more readily available and we focus on this here. 

\subsection{Correlations between galaxy photometry and halo formation redshift}

We consider the sample of IllustrisTNG halos at $z_o=1$ used for the results in Section~\ref{sec:ePStest}. For each halo in the sample we select its central galaxy, identified by the largest subhalo within the parent halo. All the galaxy properties, including $M_{*}$, SFR and photometric quantities are taken directly from the Subfind subhalo catalogs released by IllustrisTNG. We only consider star-forming (SFR$>0$) galaxies with stellar mass $M_{*}>10^8 M_{\odot}h^{-1}$.

The full sample contains galaxies with different properties, spanning over a large range of halo and stellar masses. Real galaxy samples used for observational purposes, instead, always rely on some selection criteria such as magnitude cuts, which reduce the range of variation of some properties. Therefore, besides the full sample of galaxies we also consider 8 subsamples, through which we can assess how the optimal proxy depends on the sample selection criterion.

These 8 subsamples are the following ones:
\begin{itemize}
    \item 3 subsamples  selected considering different halo mass ranges $M_h \in$ $[5\times10^{11},2\times10^{12}]$,$[2\times10^{12},10^{13}]$,$[10^{13},10^{14}]$ $M_{\odot}h^{-1}$ which we refer to as $M_h^{\,1}$,$M_h^{\,2}$,$M_h^{\,3}$ respectively;
    \item 3 subsamples obtained with stellar mass cuts $M_* \in$ $[10^{8},2\times10^{9}]$,$[2\times10^{9},3\times10^{10}]$,$[3\times10^{10},10^{12}]$ $M_{\odot}h^{-1}$ which we refer to as $M_s^{\,1}$,$M_s^{\,2}$,$M_s^{\,3}$ respectively;
    \item 2 subsamples  identified by specific star formation rate cuts, $log[sSFR]>-9.23$ and $log[sSFR]<-9.23$, respectively mimicking the populations of emission-line galaxies (ELG) and luminous red galaxies (LRG).
\end{itemize}
We only consider the heaviest galaxies within the ELG- and LRG- like subsamples, as to match the DESI target number densities for ELG, $n=7\times10^{-4}$Mpc$^{-3}h^{3}$ and LRG, $n=2\times10^{-4}$Mpc$^{-3}h^{3}$ \cite{Zhou_2023,Raichoor_2023}. These cuts follow the sample definitions adopted by \cite{sullivan_learning_2023}, based on previous studies about the link between IllustrisTNG and DESI target galaxies\cite{hadzhiyska_galaxy-halo_2021,Yuan_2022}. For each galaxy in the sample considered, we focus on the set of 8 photometric bands $X=\{U, B, V, K, g, r, i, z\}$, with AB magnitudes based on the summed-up luminosities of all the stars within the galaxy. Taking the differences between these magnitudes, we form a set of 28 colors $C=X_i - X_j$, with $X_{i,j} \in X$.

\begin{figure}[H]
 \begin{minipage}{.31\textwidth}
	\includegraphics[width=1\linewidth]{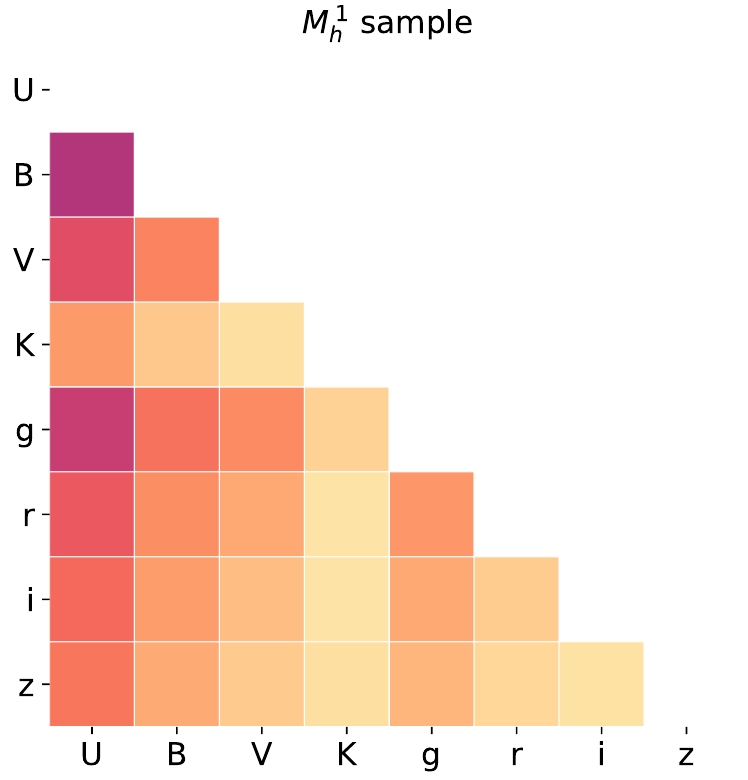} 
\end{minipage}
\begin{minipage}{.31\textwidth}
	\includegraphics[width=1\linewidth]{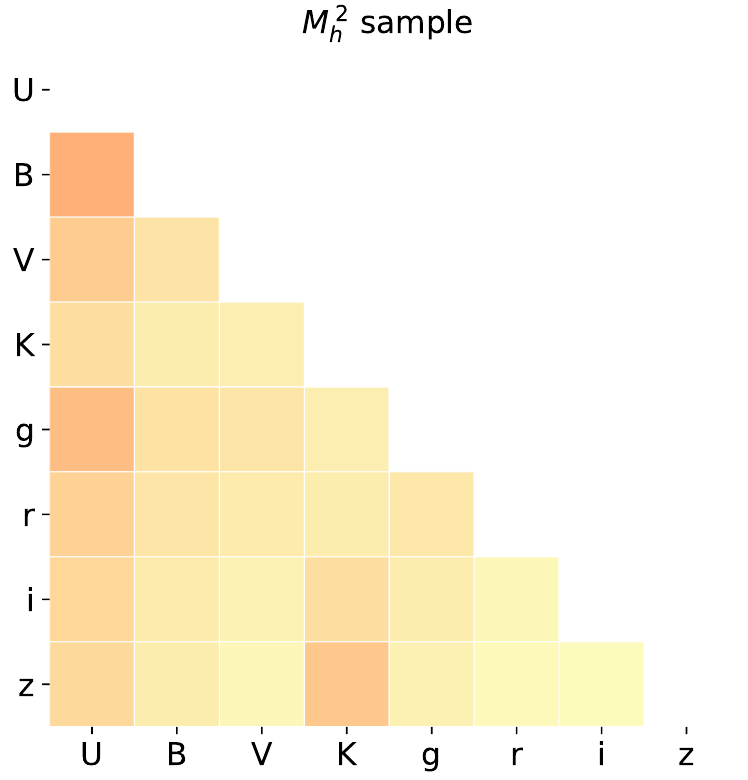} 
\end{minipage}
\begin{minipage}{.31\textwidth}
	\includegraphics[width=1\linewidth]{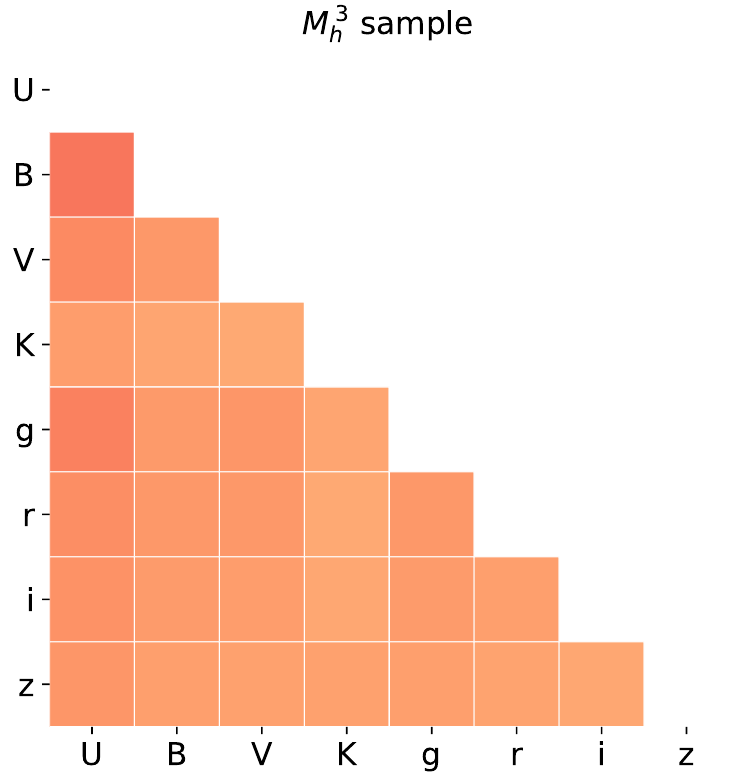}
\end{minipage}

    \vspace{0.43cm}
    
\begin{minipage}{.31\textwidth}
    \centering
	\includegraphics[width=1\linewidth]{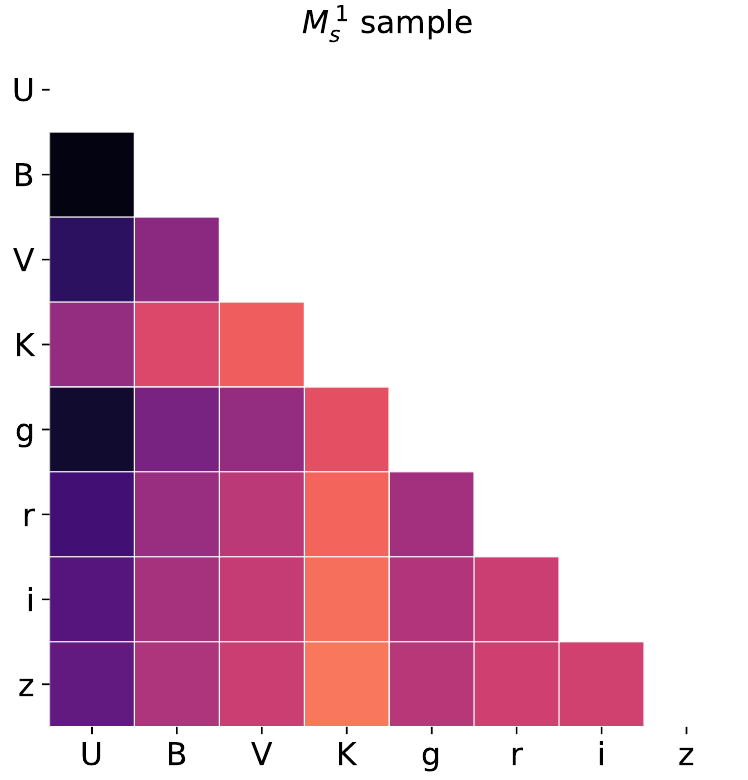} 
\end{minipage}
\begin{minipage}{.31\textwidth}
    \centering
	\includegraphics[width=1\linewidth]{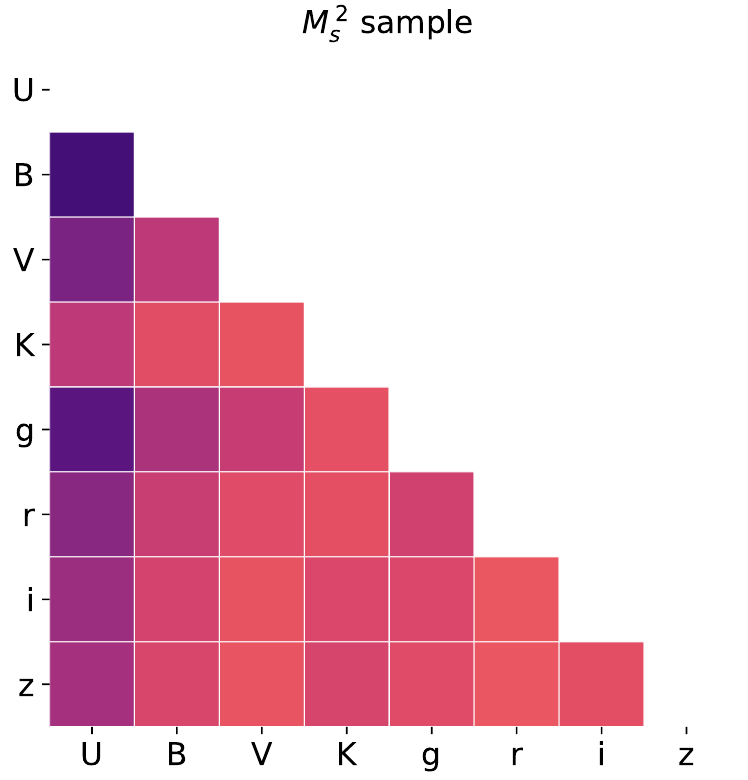} 
\end{minipage}
\begin{minipage}{.38\textwidth}
    \centering
	\includegraphics[width=1\linewidth]{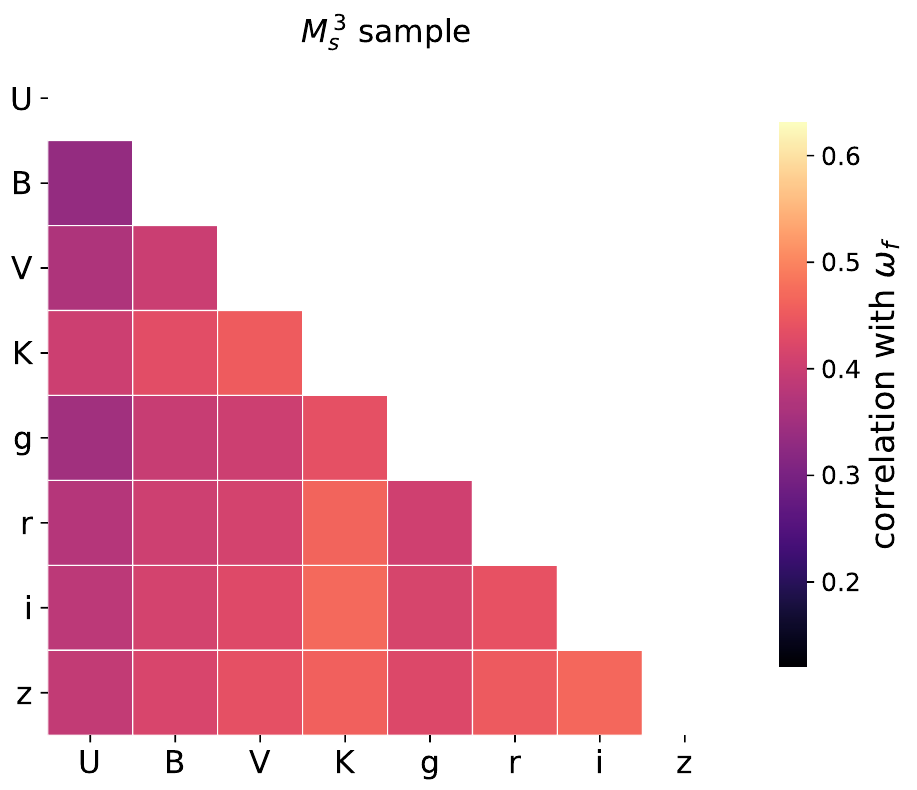}
\end{minipage}

    \vspace{0.43cm}
    
\begin{minipage}{.31\textwidth}
	\includegraphics[width=1\linewidth]{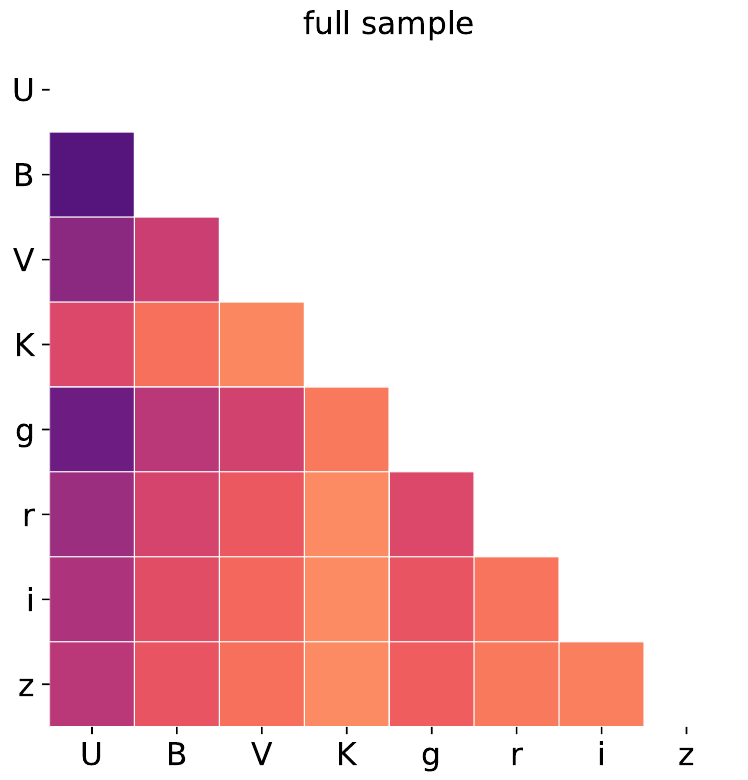}
\end{minipage}
\begin{minipage}{.31\textwidth}
	\includegraphics[width=1\linewidth]{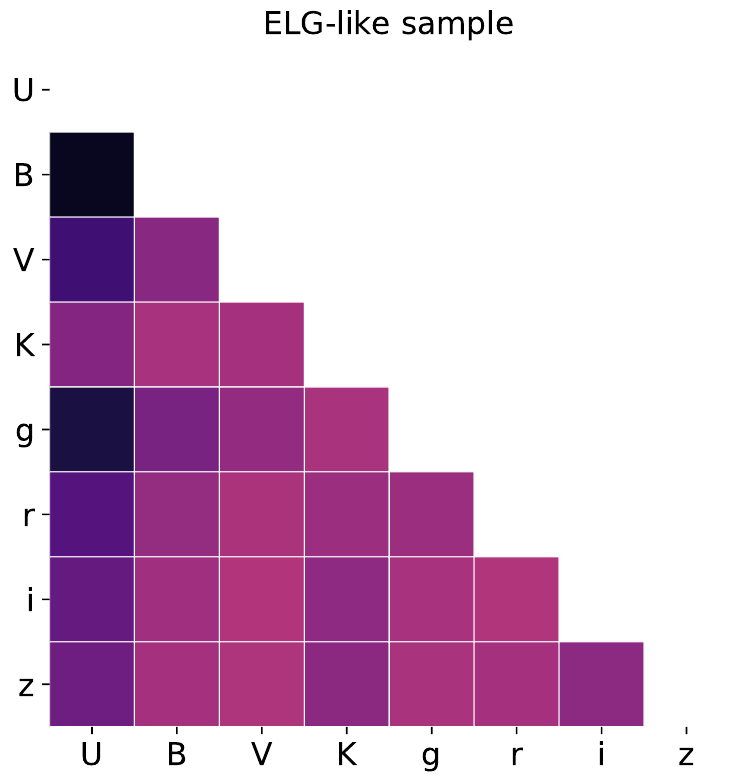} 
\end{minipage}
\begin{minipage}{.31\textwidth}
	\includegraphics[width=1\linewidth]{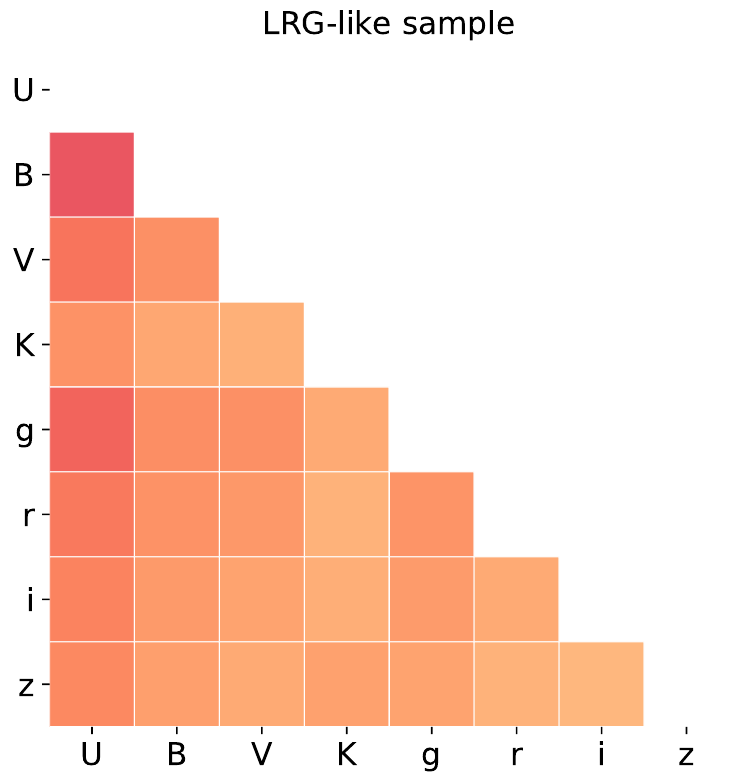}
\end{minipage}
      
\caption{Correlation between the 28 galaxy colors and $\omega_f$. Each pixel $(i,j)$ identifies a galaxy color $C=X_i - X_j$ and is color-coded depending on the correlation $r(C,\omega_f)$, ranging from dark purple for low $r$ to light yellow for high $r$. The different panels illustrate the results for different subsamples, as defined in the text.}
\label{fig:wf_corr}  
\end{figure}

As a preliminary step,  Fig.~\ref{fig:wf_corr}  illustrates the correlations between these galaxy colors and the $\omega_f$ of their host halo, for both the full sample and the 8 subsamples defined above.

The  correlation between a given galaxy color and $\omega_f$ depends on the subsample, but some common trends can be identified.
In particular, colors involving the $U$ band appear to be the least correlated with $\omega_f$, while  colors involving the $K$ band show the largest correlations.
This can be understood by taking into account that this band, centered in the infrared, can be used to track the age of a galaxy \cite{de_putter_designing_2017,bell_optical_2003}, which is in turn related to the formation time of the host halo.

Galaxy colors in the halo mass-selected subsamples manifest larger correlations, reaching values beyond $r=0.6$, so such samples would be ideal targets to estimate $\omega_f$ from photometric quantities. However, their construction is based on the knowledge of the halo mass, which needs to be evaluated indirectly, as for example by measuring the linear bias $b_1$ from the power spectrum \cite{Scoccimarro_2001,Tinker_2010}.

Subsamples more directly related to  single-object observational properties  are instead the stellar mass selected $M_s^{\,i}$ and the ELG/LRG-like ones. Galaxy colors in the $M_s^{\,i}$ samples roughly share the same behavior, although we can observe that for $M_s^{\,1}$ , the correlation varies on a wider range, compared to higher $M_s$ subsamples as $M_s^{\,3}$. In other words, for low stellar mass galaxies, we can clearly distinguish between colors that are notably good proxies for $\omega_f$ and colors that do not give much information about the halo formation time. On the other hand, for more massive galaxies there are not sharp differences among the different colors. This features can be also identified within the $M_h^{\,i}$ samples.
 
As it is evident, the LRG-like sample shares common correlation properties with the $M_h^{\,3}$ sample, since LRGs are tipically hosted by massive halos. The colors of ELGs, on the other hand do not appear to track the halo formation time.

\subsection{Optimal proxy for $\omega_f$ from a combination of galaxy colors}
\label{ssec:omegaf_proxy}
In order to select the optimal galaxy proxy of halo assembly bias, we choose a simple {\it ansatz} and assume that $\omega_f$ can be recovered from a linear combination of 3 galaxy colors $C_i$ for $i=1,2,3$, $Y=a_1 C_1+a_2 C_2+a_3 C_3$. We have verified that including more than 3 colors does not improve the results further, as discussed in the Appendix \ref{appendix}. For each possible choice of $C_1,C_2,C_3$, the coefficients $a_1,a_2,a_3$ are optimized as to maximize the Pearson correlation $r(Y,\omega_f)$. Among the 3276 possible combinations of 3 colors that can be extracted from the full set of 28 colors, the one which gives the highest $r$ is selected as optimal proxy.
This procedure is repeated for both the full sample and the 8 subsamples mentioned above, thus we obtain a different optimal proxy $Y$ for each sample.

\begin{figure}[t]
\includegraphics[width=1\linewidth]{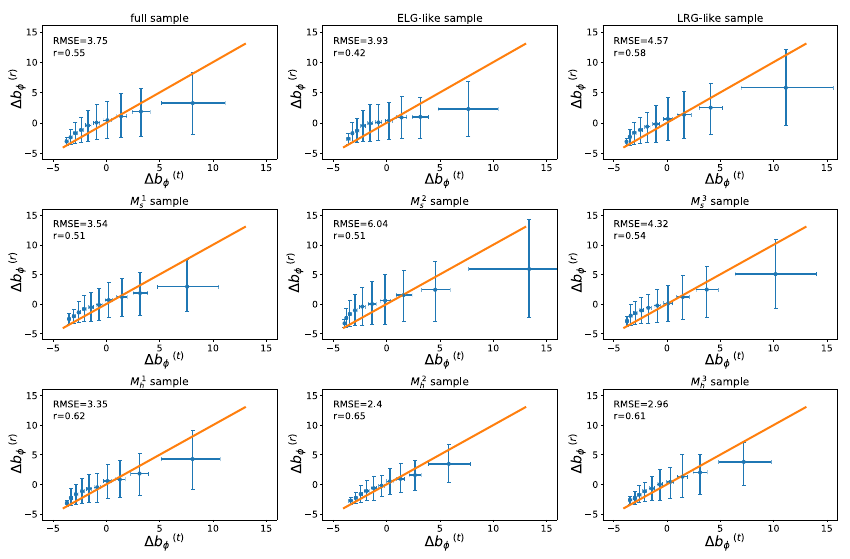}
\caption{Recovered $\Delta b_{\phi}^{(r)}$ of subsamples of galaxies selected by using the optimal galaxy proxy $Y$, as a function of their true $\Delta b_{\phi}^{(t)}$. Each point identifies a subsample, representing the 10\% quantile of the sample of galaxies (IllustrisTNG galaxies at $z_o=1$). For both the recovered and the true ones, the mean value and the 16th-84th percentiles of the $\Delta b_{\phi}$ distribution within each quantile are reported. In each panel, the correlation $r(Y,\omega_f)$ as well as the ${\rm RMSE}$ of the recovered vs true $\Delta b_{\phi}$ are shown, together with the identity relation indicated by the orange line. The different panels illustrate the results for different subsamples, as defined in the text.}
\label{fig:dbphi_proxy}  
\end{figure}

Once the optimal proxy has been determined for each sample, we test the proxy-recovered $\Delta b_{\phi}^{(r)}$ against the true $\Delta b_{\phi}^{(t)}$ (which assumes an exact knowledge of $\omega_f$, through $M_o,z_f$ of the host halo). To this end, for each sample we construct $P_{\omega_f}$, fit for $\omega_0$ in Eq.~\eqref{eq:pwf_e2ps} and then use Eq.~\eqref{eq:delta_bphi} to predict $\Delta b_{\phi}$ for each galaxy. Then, on one hand galaxies are rank ordered according to their $\omega_f$; on the other hand, according to their proxy $Y$. These two samples, containing the same galaxies with different ordering, are divided into 10\% quantiles. In the ideal case of maximum correlation $r(Y,\omega_f)=1$, the orderings of the two samples coincide, while in general there is a mismatch.

Consequently, within each $Y$-ranked quantile the distribution of $\Delta b_{\phi}^{(r)}$ will differ from that of $\Delta b_{\phi}^{(t)}$, related to the corresponding $\omega_f$-ranked quantile. We evaluate this mismatch by considering the root mean square error ${\rm RMSE}=\sqrt{\displaystyle\sum_{n}^{N} \frac{{(\Delta b_{\phi}^{(r)}-\Delta b_{\phi}^{(t)})}^2}{N}}$, with $N$ the size of the galaxy sample, that quantifies the scatter around the identity relation.

In Fig.~\ref{fig:dbphi_proxy} we report the mean value and 16th-84th percentiles of the $\Delta b_{\phi}^{(r)}$ and $\Delta b_{\phi}^{(t)}$ distributions within the 10 quantiles, one as a function of the other, for all the samples considered in the analysis.

Combining galaxy colors together allows one to get larger correlations with $\omega_f$, up to $r=0.65$. The different ``performance" of the various samples (i.e. the maximum correlation between galaxy colors and $\omega_f$) mirrors the results in Fig.~ \ref{fig:wf_corr}, with the halo mass selected samples manifesting largest correlations and ELG-like one, the smallest ones. Generally, larger $r$ does not guarantee to have a lower ${\rm RMSE}$. In other terms, a good prediction of $\omega_f$ through the proxy $Y$ does not ensure to recover $\Delta b_{\phi}$ optimally. This is due to the properties of the $\omega_f$ distribution within each sample and the nonlinear relation between $\omega_f$ and $\Delta b_{\phi}$.

More specifically, as one can see in Fig.~\ref{fig:eps_test}, $\Delta b_{\phi}$ varies rapidly for large $\omega_f$, while it stays roughly constant for small $\omega_f$. This propagates into a larger dispersion in both $\Delta b_{\phi}^{(t)}$ and $\Delta b_{\phi}^{(r)}$ within the high $\omega_f$ quantiles. As a consequence, samples containing old halos are affected, as the LRG-like one, which has a relatively large ${\rm RMSE}$, although $r=0.58$.

In general, although most of the proxies have good correlations with $\omega_f$, our simple model is not flexible enough to fully predict $\omega_f$ and recover $\Delta b_{\phi}^{(t)}$ completely. Specifically, for low (high) $\omega_f$ quantiles, $\Delta b_{\phi}^{(r)}$ is always over-(under-)estimated with respect to $\Delta b_{\phi}^{(t)}$. We will return to this systematic shift in Sec.~\ref{ssec:singletracer}. On the one hand, predictions may be improved by considering nonlinear models or using machine learning techniques, possibly adding further information beyond galaxy colors; this is left for future work. On the other hand, simple models have the advantage of being more robust to changes in simulations settings and parameters, as will be shown in Sec.~\ref{sec:camels}.

\section{Marginalizing over cosmology and astrophysics: tests on CAMELS-TNG}
\label{sec:camels}

Galaxy properties are determined by complex physical aspects, often referred to as baryonic effects, which also affect the underlying dark matter distribution through feedback processes. These include feedback from two main sources: active galactic nuclei (AGNs) and starbursts, the latter determined by supernova explosions and stellar winds ejected from newborn stars. 

Describing these processes in cosmological simulations is very challenging: the scales involved are beyond the resolution of large cosmological boxes and phenomenological models are usually adopted, with parameters chosen as to match observations. The details of their implementation in IllustrisTNG can be found in \cite{Pillepich_2017}. Specific simulations make specific implementation choices, but there is an intrinsic theoretical uncertainty associated. In Section~\ref{sec:proxy} the assembly bias proxy has been calibrated for the specific implementation of IllustrisTNG; calibration on a different phenomenological prescription might result in a different proxy. 

In order to account for the theoretical uncertainty associated to these phenomenological models, an approach is to marginalize over the parameters which define them. This is one of the purposes of the suite of CAMELS simulations \cite{villaescusa-navarro_camels_2021}, which scan different subgrid models for galaxy formation and various choices of astrophysical and cosmological parameters. Among the different models considered in CAMELS, we focus here on the subset of CAMELS-TNG simulations, run with the same subgrid prescription of IllustrisTNG. Although tests on other subsets as CAMELS-SIMBA and CAMELS-ASTRID\cite{ni2023camels} would provide a broader comparison among different models, galaxy photometry is not yet available for these sets.

The set of CAMELS-TNG simulations includes 1000 different simulations run with different initial random seed and cosmological and astrophysical parameters ranging in a Latin Hypercube with $\Omega_m\in[0.1, 0.5]$, $\sigma_8\in[0.6, 1.0]$, $A_{\rm SN1}\in[0.25, 4.0]$, $A_{\rm SN2}\in[0.5, 2.0]$, $A_{\rm AGN1}\in[0.25, 4.0]$, $A_{\rm AGN2}\in[0.5, 2.0]$. The 4 astrophysical parameters regulate starbursts and AGN feedback and are set to 1 in IllustrisTNG.
CAMELS implements a resolution comparable with that of the original TNG300, while simulating a smaller box with $L=25$ Mpc $h^{-1}$. We refer the reader to \cite{villaescusa-navarro_camels_2021} for further details on the simulations.

Our goal is to use CAMELS-TNG galaxies to assess the robustness of the proxy calibration of Section \ref{sec:proxy}, which is tuned to the specific choice of astrophysical (and cosmological) parameters in IllustrisTNG.
We analyse the snapshots at $z_o=0.95$ of each simulation. For each simulation, we consider the photometric data of the central galaxies and compute $\omega_f$ of their host halo, which depends on cosmology via $D(z)$ and $\sigma (M)$ appearing in Eq.~\eqref{eq:omega_f}. All the galaxies belonging to different simulations (i.e., formed under different astrophysical models depending on the values of the baryonic feedback parameters) are put together in what we refer to as “full sample”. By analysing this sample we are effectively marginalizing over cosmology, (IllustrisTNG) astrophysics and including the effect of cosmic variance.

The different simulation settings in CAMELS give rise to differences with the IllustrisTNG galaxy sample used in Section \ref{sec:proxy}. In particular, the smaller size of the simulation box has an impact on the most massive galaxies which can be simulated, making the selection of an LRG-like sample impossible. Therefore we restrict our analysis to the full sample.

\begin{figure}[t]
    \begin{minipage}{.33\textwidth}
    \centering
	\includegraphics[width=1\linewidth]{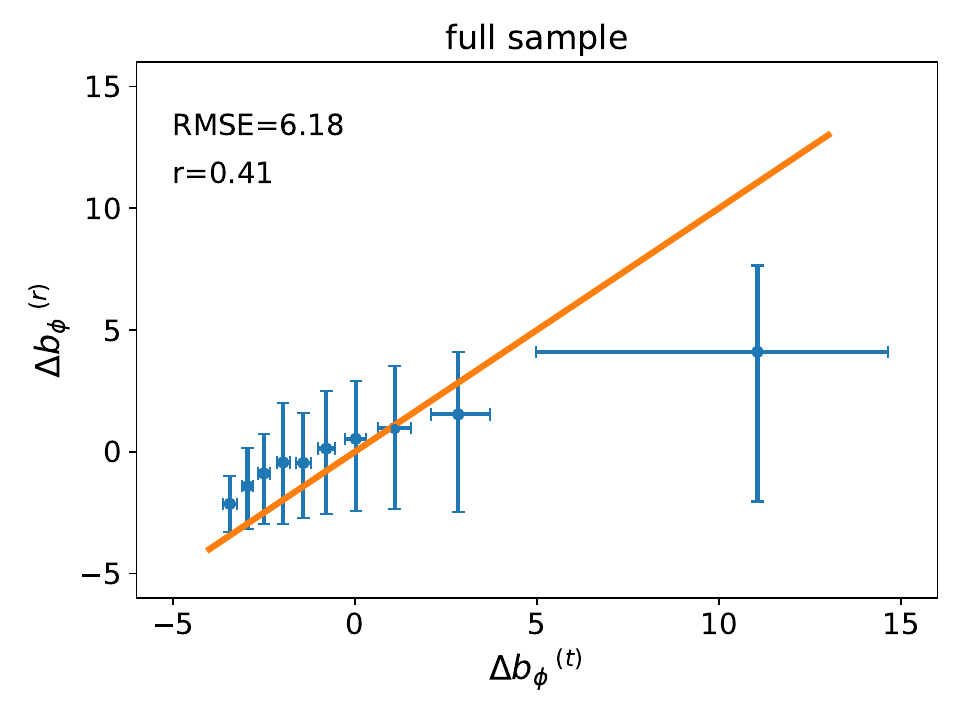} 
    \end{minipage}
    \begin{minipage}{.33\textwidth}
    \centering
	\includegraphics[width=1\linewidth]{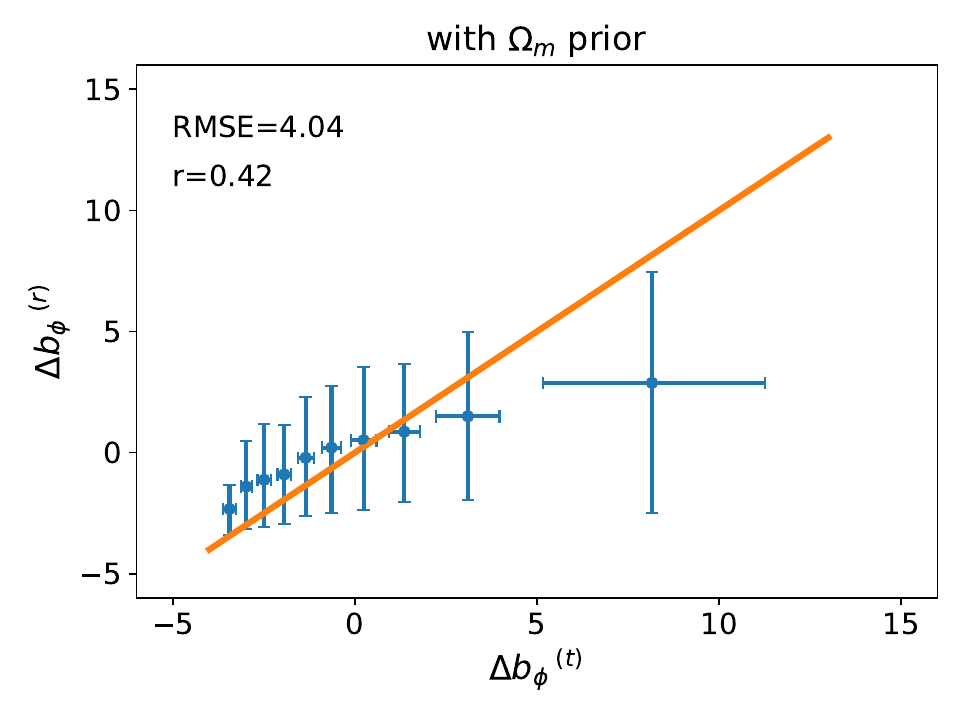} 
    \end{minipage}
    \begin{minipage}{.33\textwidth}
    \centering
	\includegraphics[width=1\linewidth]{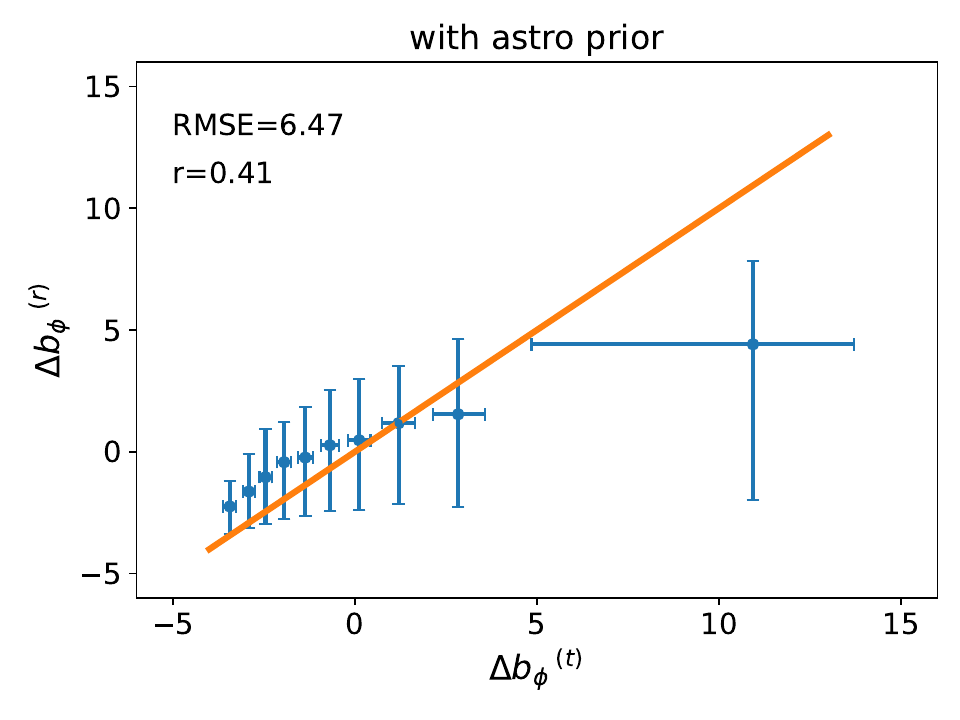} 
    \end{minipage}
    \begin{minipage}{.33\textwidth}
    \centering
	\includegraphics[width=1\linewidth]{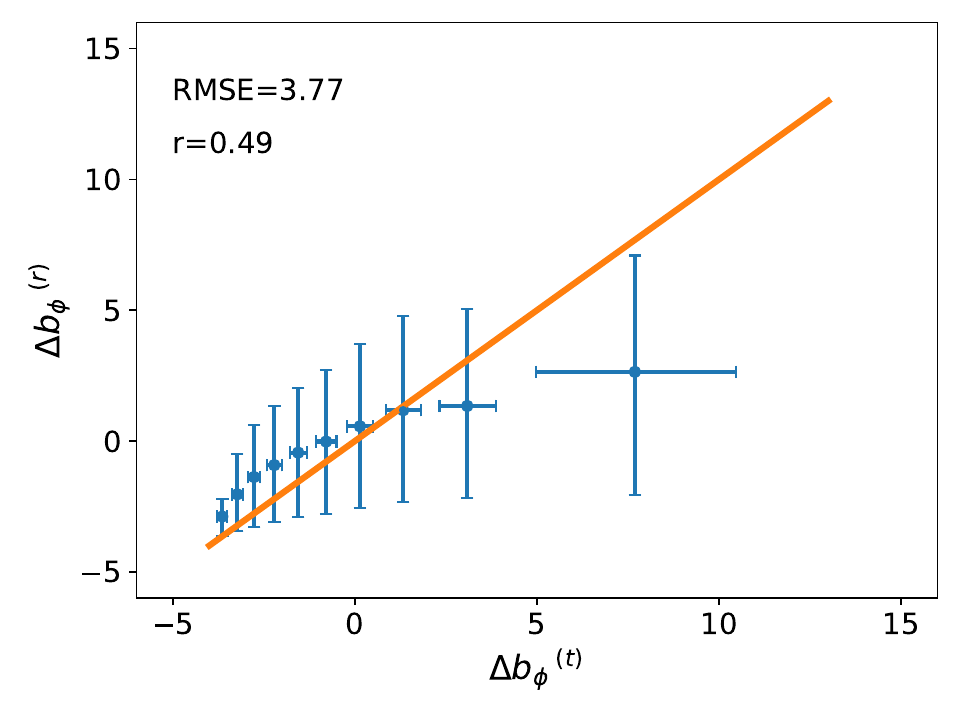} 
    \end{minipage}
    \begin{minipage}{.33\textwidth}
    \centering
	\includegraphics[width=1\linewidth]{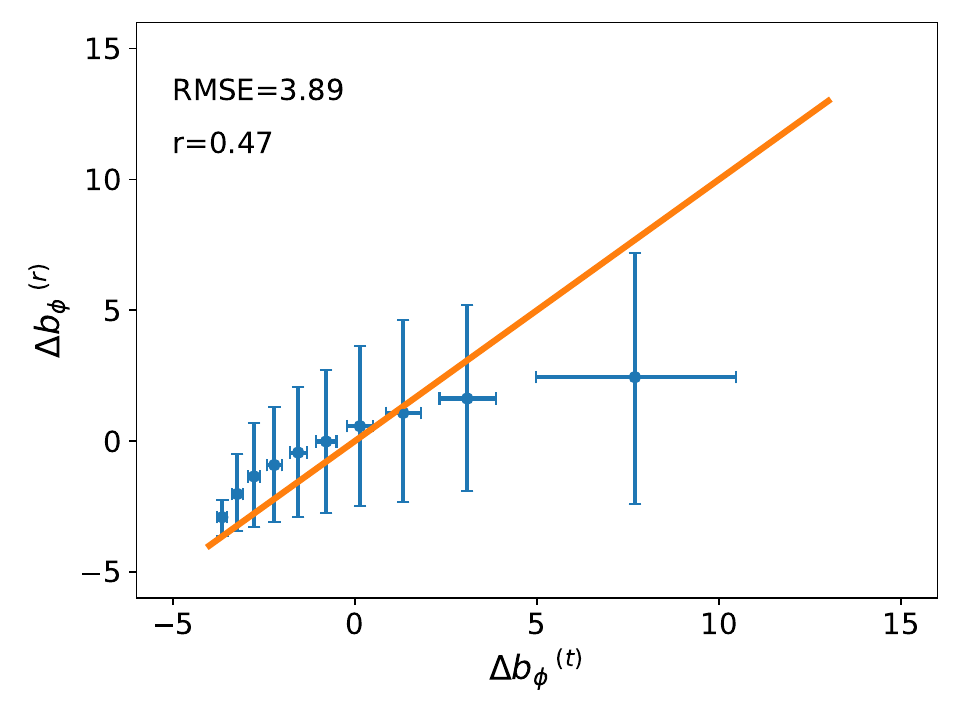} 
    \end{minipage}
    \begin{minipage}{.33\textwidth}
    \centering
	\includegraphics[width=1\linewidth]{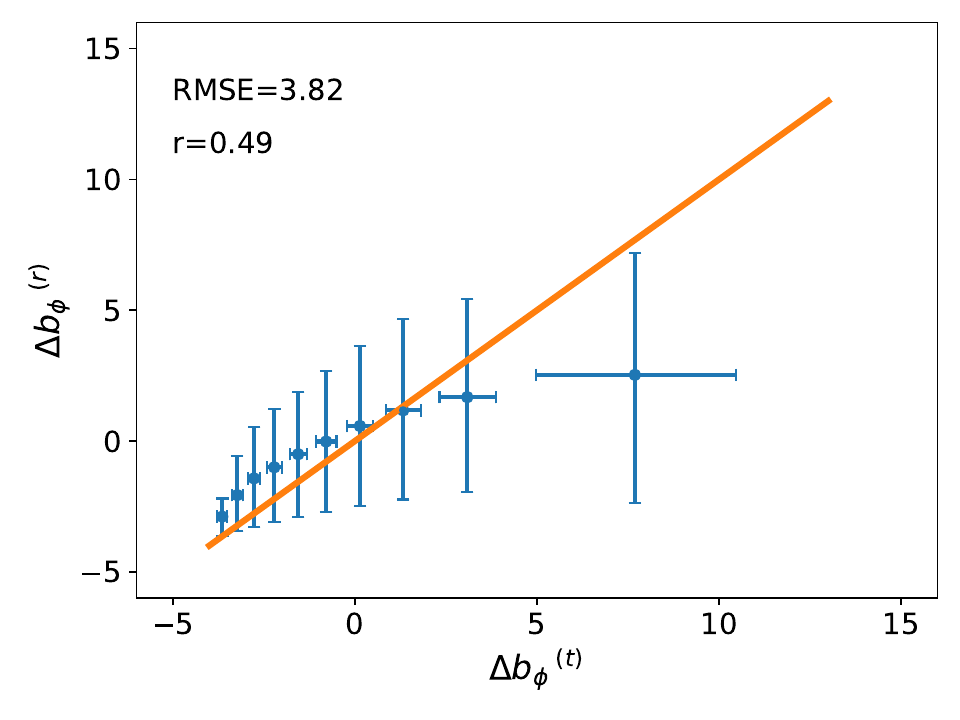} 
    \end{minipage}
\caption{Recovered $\Delta b_{\phi}^{(r)}$ of subsamples of galaxies selected by using the optimal galaxy proxy $Y$, as a function of their true $\Delta b_{\phi}^{(t)}$, similarly to Fig.~\ref{fig:dbphi_proxy}. In the top panels, we report the results of the calibration of the proxy on CAMELS. Both the full sample and samples where either an $\Omega_m$ prior or a prior on  astrophysical parameters are considered. In the bottom panels, the CAMELS-calibrated proxy is applied to IllustrisTNG full sample, in comparison with the first panel of Fig.~\ref{fig:dbphi_proxy}.}
\label{fig:dbphi_proxy_camels}  
\end{figure}

By using the same method described in Section \ref{sec:proxy}, we calibrate the optimal proxy on CAMELS-TNG full sample and then test it on IllustrisTNG full sample. The results are reported in the left column of Fig.~\ref{fig:dbphi_proxy_camels}, where the top panel is related to the calibration on CAMELS-TNG sample and the bottom panel to the test on IllustrisTNG sample. 

Not unexpectedly, the proxy performance in the calibration sample degrades, effect that is driven in part by the variation in cosmologies and in part by the variation of the  astrophysical parameters. In particular, marginalizing over $\Omega_m$ has a large impact on $P_{\omega_f}$. Indeed, at fixed formation redshift $z_f$, extreme values of this parameter can produce very different values of $\omega_f$ (see Eq.~\eqref{eq:omega_f}), broadening the  distribution $P_{\omega_f}$. This larger scatter in $\omega_f$ is propagated to $\Delta b_{\phi}$ differently for low and high $\omega_f$ values, as we can notice from Fig.~\ref{fig:eps_test}. In particular, as $\Delta b_{\phi}$ grows rapidly for older halos, the $\Delta b_{\phi}^{(t)}$ distribution of the last 10\% quantile gets skewed towards larger values with respect to the IllustrisTNG counterpart, illustrated in the bottom left panel. The distribution of the recovered $\Delta b_{\phi}^{(r)}$ is also affected by the marginalization over the astrophysical parameters, which mixes the relations between $\omega_f$ and the galaxy colors.

However, the test on IllustrisTNG of the CAMELS-calibrated proxy shows good results if compared with the IllustrisTNG-calibrated one (illustrated in the first panel of Fig.~\ref{fig:dbphi_proxy}). The correlation with $\omega_f$ only lowers from $r=0.55$ to $r=0.49$ and the scatter around equality is practically unchanged, from ${\rm RMSE}=3.75$ to ${\rm RMSE}=3.77$. Therefore, despite the large range of variation of the CAMELS galaxy features, due to the marginalization over cosmological and astrophysical parameters, the performance of a CAMELS-calibrated proxy is marginally affected.

To partly disentangle the two effects (marginalization over cosmology vs marginalization over astrophysics) we repeat the analysis by imposing either a cosmology prior or priors on the astrophysical parameters.

The cosmology prior is imposed by considering the subset of 500 simulations that exclude extreme values of $\Omega_m$, which has been tested to have a larger impact on $P_{\omega_f}$ with respect to $\sigma_8$. Specifically, our prior reduces its range of variation to $\Omega_m\in[0.2, 0.4]$, symmetrically around the fiducial value of $0.3$. The range of variation of the astrophysical parameters is instead left unchanged. The results about the calibration and the test are shown in the middle column of Fig.~\ref{fig:dbphi_proxy_camels}.

The astrophysical parameters prior is imposed by considering the subset of 500 simulations that are closer to the IllustrisTNG fiducial values. The ranges of variation are $A_{\rm SN1}\in[0.3, 3.0]$, $A_{\rm SN2}\in[0.58, 1.7]$, $A_{\rm AGN1}\in[0.3, 3.0]$, $A_{\rm AGN2}\in[0.58, 1.7]$ and the results are in Fig.~\ref{fig:dbphi_proxy_camels}, on the right.

Within the calibration sample, imposing a cosmology prior improves the correlation coefficient and reduces the scatter, while imposing an astrophysical parameters prior does not. As regards the test on IllustrisTNG, both proxies have a performance roughly as good as the one calibrated on the full sample.

In Table \ref{tab:proxies} we summarize the combinations which define the optimal proxies, as obtained from the analyses in Sections \ref{sec:proxy}, \ref{sec:camels}.

\begin{table}[t]
    \centering
    \begin{tabular}{c|c|c}
    calibration sample & proxy $Y$ & $r(Y,\omega_f)$\\
    \hline \hline
        LRG-like & $6.3(V-K) + 3.1(B-r) + 0.005(i-U)$ & 0.58\\
        ELG-like & $4(V-r)+2(i-g)+0.8(g-z)$ & 0.42\\
        IllustrisTNG full & $3.8(V-K)+1.8(B-z)+0.01(K-U)$ & 0.55\\
        CAMELS full & $12.4(V-i)+1.9(K-B)+(r-U)$ & 0.41\\
        $M_h^{\,1}$ & $4(i-r)+2.5(V-z)+0.2(z-U)$ & 0.62\\
        $M_h^{\,2}$ & $2(V-g)+(B-r)+0.02(V-U)$ & 0.65\\
        $M_h^{\,3}$ & $2(V-g)+(B-r)+0.014(z-U)$ & 0.61\\
        $M_s^{\,1}$ & $8.25(g-B)+1.02(i-U)+(K-r)$ & 0.51\\
        $M_s^{\,2}$ & $10(V-r)+3(V-B)+0.2(K-z)$ & 0.51\\
        $M_s^{\,3}$ & $1.96(V-B)+1.73(B-g)+(K-r)$ & 0.54

    \end{tabular}
    \caption{Summary of the combinations of galaxy colors obtained for the optimal proxies calibrated on the respective sample, reported in the first column. The optimal proxy is reported in  the second column and  the third column reports the Pearson correlation coefficient between the proxy and the physical variable $\omega_f$ (see text for more details). }
    \label{tab:proxies}
\end{table}

\section{Forecasts on $\fnl$ measurement}
\label{sec:forecasts}

The observational proxies calibrated in Sections~\ref{sec:proxy} and \ref{sec:camels} provide an inference for $b_{\phi}$ for each galaxy sample, which can be used as a prior when inferring constraints on $f_{\rm NL}$.
Although a prior knowledge of $b_{\phi}$ is not needed to \textit{detect} local PNG, it is required to \textit{constrain} the value of $\fnl$. 
The precision and accuracy of the $b_{\phi}$ prior will affect the error of the $\fnl$ measurement in two ways: statistical, $\sigma_{\fnl}$, and systematic $\sigma^{(sys.)}_{f_{\rm NL}}$. The statistical error achievable on $f_{\rm NL}$ depends on the bias $b_{\phi}$ of the sample because the noise in the data is fixed effectively, for small non-Gaussianity, by the power spectrum amplitude, the volume surveyed and the shot noise  but  the signal itself for a given $f_{\rm NL}$ is proportional to $b_{\phi}$. Thus samples with larger  $|b_{\phi}|$ yield better signal to noise in the data and thus smaller errors on $f_{\rm NL}$. A sample that spans a range of $\omega_f$ will  tend to have reduced $|b_{\phi}|$ compared to a sample with well selected $\omega_f$. 

On the other hand, a systematic error in the estimate of $b_{\phi}$ of $\sigma^{(sys.)}_{b_{\phi}}$ will introduce a systematic error in $f_{\rm NL}$ of $\sigma^{(sys.)}_{f_{\rm NL}}= \sigma^{(sys.)}_{b_{\phi}} b_{\phi}^{-1}f_{\rm NL}$, biasing the $\fnl$ constraint. In what follows, we report the results of the analysis of these two different contributions to the forecasted error on $\fnl$.

The statistical error can be estimated from the Fisher information matrix as $\sigma_{\fnl}=1 / \sqrt{\mathcal{F}_{\fnl}}$, where $\mathcal{F}_{\fnl}$ reads
\begin{equation}\label{eq:fisher}
\mathcal{F}_{\fnl}=\int_{k_{min}}^{k_{max}} dk \frac{\partial {\bf D}^T (k)}{\partial \fnl}{\bf Cov}^{-1}(k)\frac{\partial {\bf D} (k)}{\partial \fnl},
\end{equation}
with {\bf D} being the data vector (i.e. the power spectrum) and {\bf Cov} the covariance matrix of the data which we assume to be diagonal. The reason is that the scale-dependent bias plays a role on large scales, where non-linearities sourced by gravitational evolution are negligible and different Fourier modes evolve independently. Moreover we do not explore here possible degeneracies between $f_{\rm NL}$ and other cosmological (or astrophysical) parameters.

In order to  provide somewhat realistic forecasts, in this Section we will consider the IllustrisTNG full sample, as well as the ELG- and LRG-like subsamples, as defined in Section \ref{sec:proxy}. For completeness, we also compare the results of the CAMELS- and the IllustrisTNG-calibrated proxies applied to the full sample. We provide both single tracer and multitracer results and refer the reader to \cite{carbone_non-gaussian_2008,barreira_can_2022,karagiannis_constraining_2018} for the details on the derivation of the equations we report in what follows.
\subsection{Single tracer: statistical error}
Let us first consider a data vector $D(k)=P_{\rm gg}(k)$ consisting in the power spectrum measurements of a single sample of galaxies at redshift $z_o$ with number density $\bar{n}$, where $P_{\rm gg}(k)$ is given by
\begin{equation}
P_{\rm gg}(k)=P_{\rm th}(k)+P_{\epsilon}\equiv\left[b_1 + b_{\phi}\fnl{\cal M}^{-1}(k,z_o)\right]^2 P_{\rm lin}(k) + \frac{1}{\bar{n}}
\end{equation}
and we have assumed Poisson shot noise. By using Eq.~\ref{eq:fisher}, the $\fnl$ error Fisher forecast becomes \cite{carbone_non-gaussian_2008,barreira_towards_2023}
\begin{equation}\label{eq:sigmafnl}
    \sigma_{\fnl}(k) = \sqrt{\frac{2(2\pi)^3}{V_SV_{k}}} \frac{b_1 + b_{\phi} f_{\rm NL} {\cal M}^{-1}(k,z_o)}{2 b_{\phi}{\cal M}^{-1}(k,z_o)} \left(1 + \frac{1}{\bar{n}P_{\rm th}(k)}\right),
\end{equation}
where $V_S$ is the volume of the survey and $V_{k}=4\pi k^2 \Delta k$ is the volume in Fourier space of the power spectrum shell at wavenumber $k$, with wavenumber bin size $\Delta k$. In order to provide quantitative results, we consider a  realistic  case  with galaxy samples to have   the same empirically calibrated DESI linear bias \cite{desi_collaboration_desi_2016}, $b_1=1.38, 1.79$ respectively for ELG- and LRG-like samples, computed at $z_o=1$. The nominal number densities of the two samples are $\bar{n}=7\times10^{-4}$Mpc$^{-3}h^{3}$ for ELG and $\bar{n}=2\times10^{-4}$Mpc$^{-3}h^{3}$ for LRG. For the full sample, we use the same linear bias and number density as the ELG-like sample. We adopt the DESI $k_{min}=0.0023\, h$ Mpc$^{-1}$ as reported by \cite{sullivan_learning_2023} and the corresponding survey volume is assumed to be $V_S=(2\pi/k_{min})^3=20.4$ Gpc$^3/h^3$.

\begin{figure}[H]
    \begin{minipage}{.33\textwidth}
	\includegraphics[width=1\linewidth]{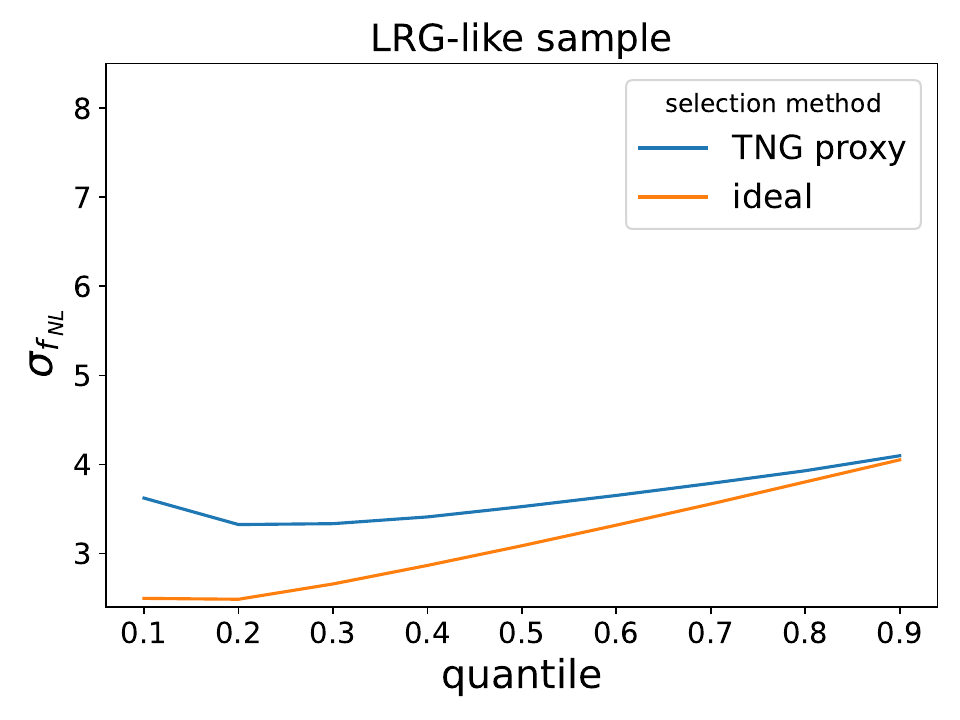}
    \end{minipage}
    \begin{minipage}{.33\textwidth}
	\includegraphics[width=1\linewidth]{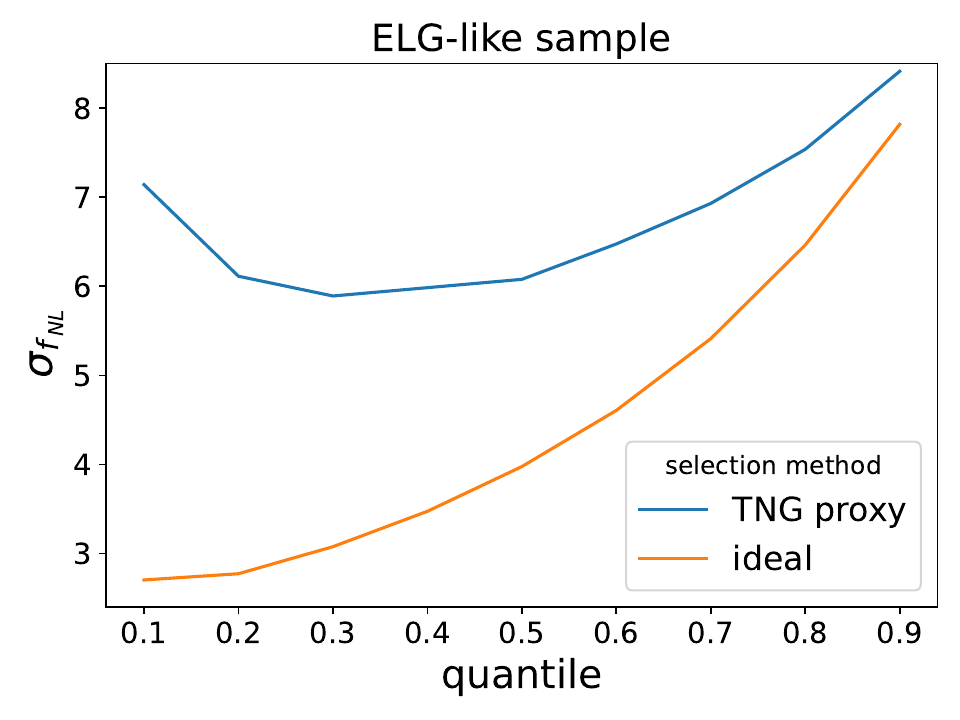} 
    \end{minipage}
    \begin{minipage}{.33\textwidth}
	\includegraphics[width=1\linewidth]{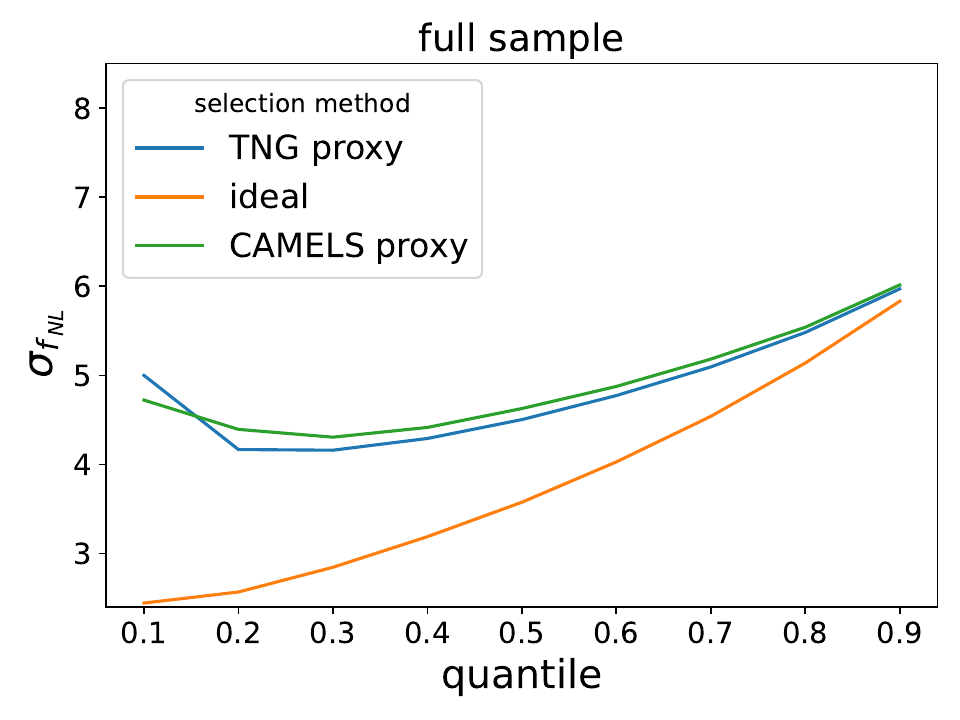} 
    \end{minipage}

	\caption{Forecasted single-tracer error $\sigma_{\fnl}$ for the ELG-like, LRG-like subsamples and the full sample. In all panels the cumulative quantiles (on the x axis) are obtained with the true $\omega_f$ for the ideal case and the proxy-selected $\omega_f$ for the proxy case. Only cumulative quantiles of the oldest (largest $\omega_f$) halos are considered. For larger quantile fractions, increasingly younger halos are included in the sample. The green line in the right panel is obtained using the CAMELS-calibrated proxy  applied to the IllustrisTNG galaxies. 
	\label{fig:sigmafnl_single}}
\end{figure}

From Eq.~\eqref{eq:sigmafnl}, we can see that for a given number density $\bar{n}$, $\sigma_{\fnl}$ decreases with increasing $b_{\phi}$ due to both the denominator and the $P_{\rm th}(k)$ in the bracket, which grows with $b_{\phi}^2$. By selecting a sample of galaxies with large $\omega_f$, hosted by the oldest halos, we can maximize $b_{\phi}$ and obtain the most precise constraints on $\fnl$. The smaller the sample, the larger the $b_{\phi}$. However, for small samples the number density $\bar{n}$ effectively introduces a tradeoff, due to the enhanced shot noise, which worsens the precision, increasing $\sigma_{\fnl}$. This is illustrated in Fig.~\ref{fig:sigmafnl_single}, where we report $\sigma_{\fnl}$ as a function of the cumulative quantile of oldest halos, for both the ideal (where we exactly know $\omega_f$ for each halo) and the proxy selected samples.
Selecting a subsample of galaxies with a large value of $b_{\phi}$ leads to better constraints than those obtained from the full sample for a volume of $20.4$ Gpc$^3 h^3$. The improvement, in the ideal case ($\omega_f$ perfectly known), becomes almost a factor of 2 for LRG and a factor of 3 for ELG. This difference between the two is due to the fact that the $\omega_f$ distribution in the ELG sample is broader, so there is more room for improvement than in the LRG case. The same conclusion applies to the full sample, in which case the improvement is almost a factor of 3.

In the realistic case, in which $20-30\%$ of galaxies most-likely hosted by older halos is selected by using a proxy as discussed above, one can enhance the precision on the $\fnl$ constraint by $20-25\%$. For the ELG-like sample, this is only a fraction of the potential improvement, due to the fact that the proxy is not optimal, as discussed in Section \ref{sec:proxy}.
Therefore, within the context of a single-tracer analysis, the LRG-like sample appears to be a better target to provide $\fnl$ constraints, with the minimum $\sigma_{\fnl}$ being almost half of the ELG-like one. Certainly, an improved calibration of the proxy in the ELG-like sample would produce better results, but we leave this to future work.
As regards the full sample, we observe that the CAMELS- and the IllustrisTNG-calibrated proxies have practically the same performance, as expected from the discussion in Section \ref{sec:camels}. The improvement in this case is somewhat in-between the LRG- and the ELG-like sample related ones.

The upturn at low quantiles in  Fig. \ref{fig:sigmafnl_single} is related to the ratio between the shot noise $1/\bar{n}$ and $P_{th}$, contributing to $\sigma_{\fnl}$ in Eq. \eqref{eq:sigmafnl}. If the subsample selected is too small, i.e. $\lesssim 20\%$, the larger shot noise dominates the error $\sigma_{\fnl}$, which increases. This effect does not show up in the ideal case, where maximizing $b_{\phi}$ determines a larger $P_{th}$ which compensates the shot noise contribution.

As Fig.~\ref{fig:sigmafnl_single} shows,  the use of the proxy only degrades the $f_{\rm NL}$ statistical constraints by a factor 1.5 or less  for LRG,  by a factor 3 or less for ELG, and by a factor of 2 or less for the full sample, compared to the ideal case where the galaxy sample can be selected by $\omega_f$ perfectly.

\subsection{Single-tracer systematic error}
\label{ssec:singletracer}
As Figs.~\ref{fig:dbphi_proxy} and~\ref{fig:dbphi_proxy_camels} show, there is a residual systematic trend in $\Delta b_{\phi}^{(r)}$ with respect to $\Delta b_{\phi}^{(t)}$. These residuals, if uncorrected, would be responsible for introducing a systematic shift in the inferred $f_{\rm NL}$. 

\begin{figure}[H]
\begin{minipage}{.45\textwidth}
	\includegraphics[width=1\linewidth]{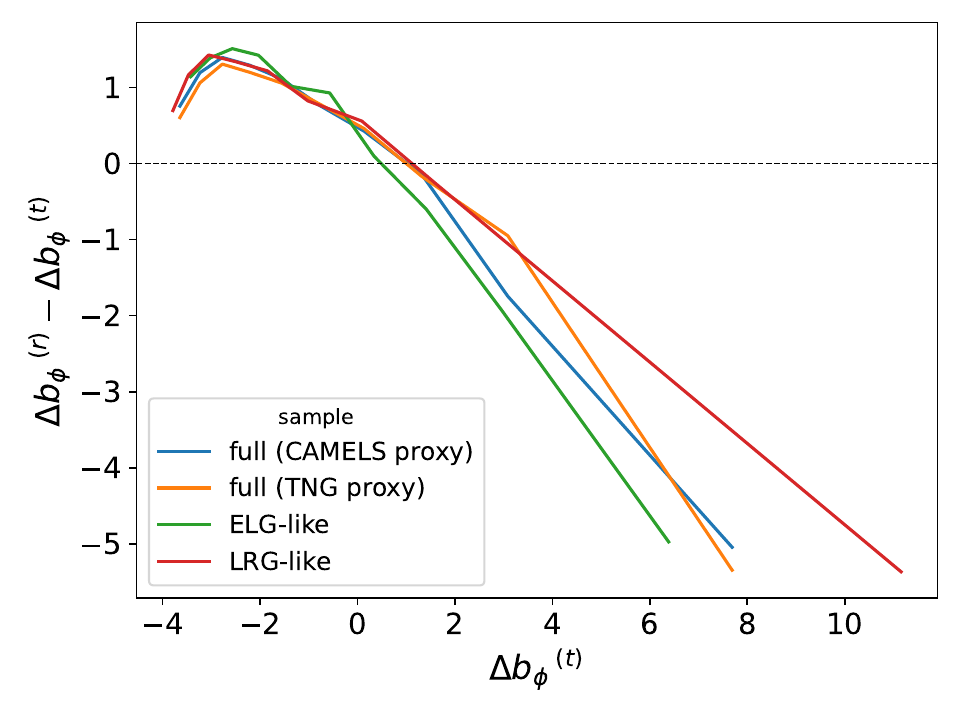}
    \end{minipage}
\hspace{0.5cm}
\begin{minipage}{.44\textwidth}
	\includegraphics[width=1\linewidth]{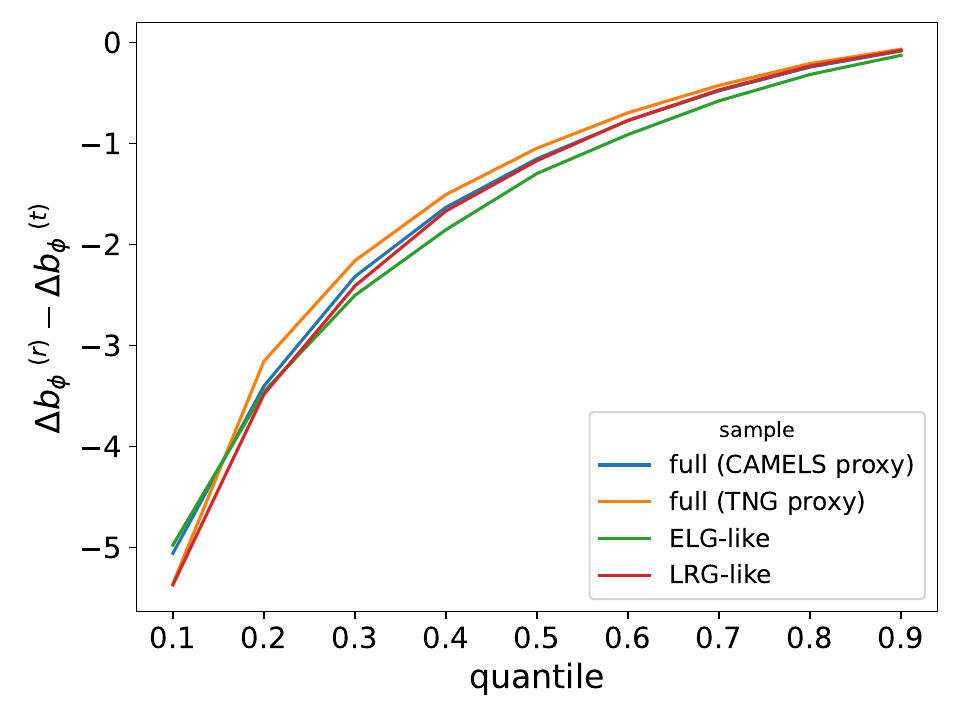}
    \end{minipage}

\begin{minipage}{.45\textwidth}
	\includegraphics[width=1\linewidth]{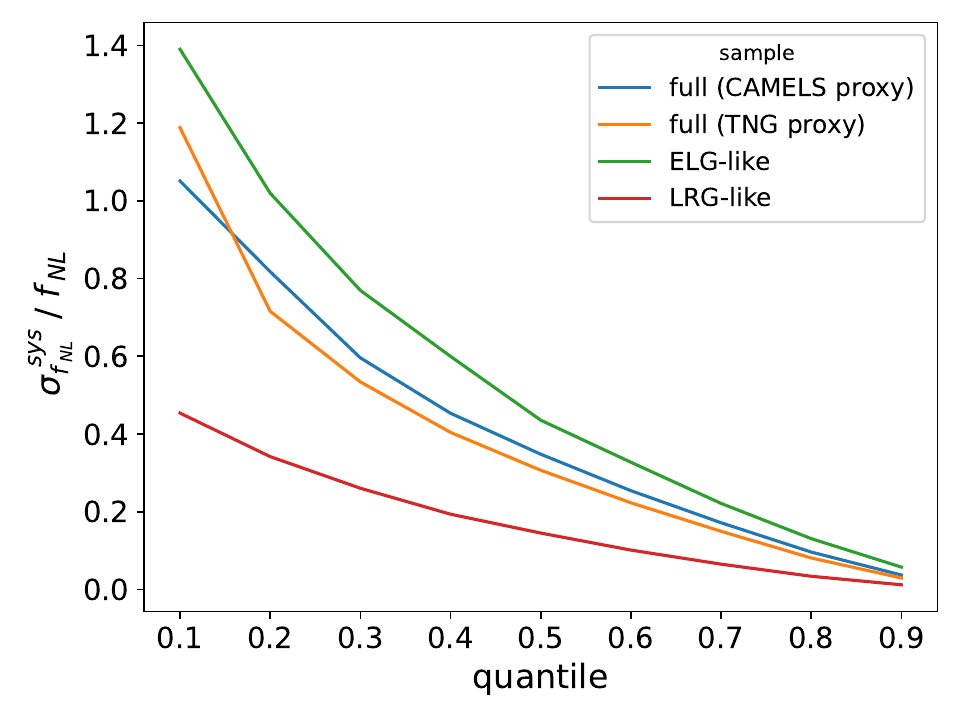}
    \end{minipage}
\hspace{0.4cm}
    \begin{minipage}{.45\textwidth}
	\includegraphics[width=1\linewidth]{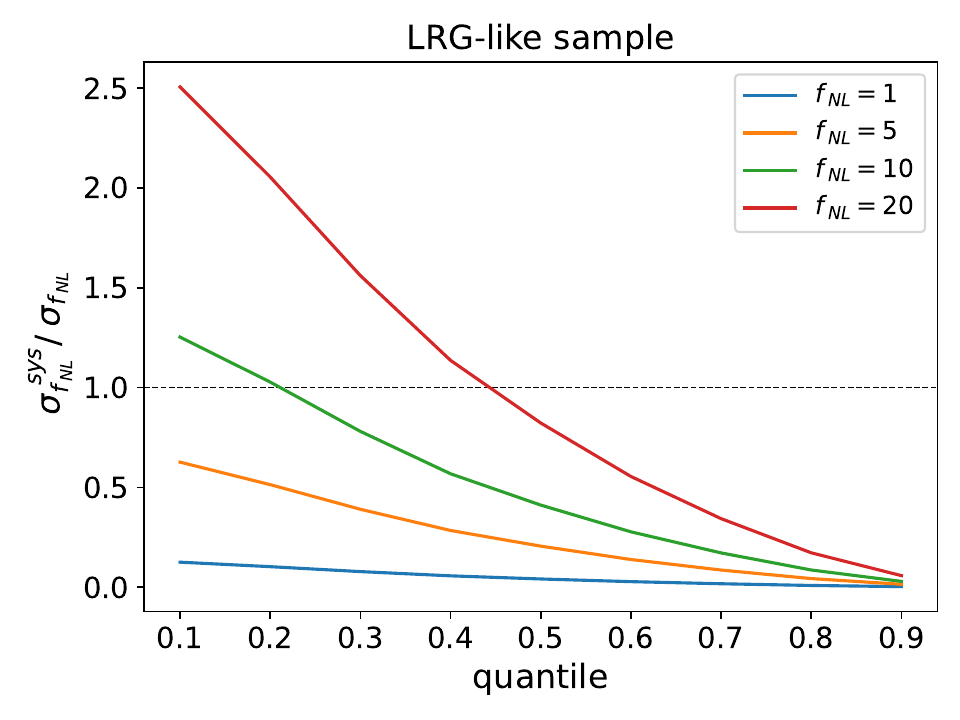} 
    \end{minipage}
\caption{Residuals and relative conservative systematic error on $\fnl$ due to the limitations of the  adopted proxy  which induces a systematic shift in the $\Delta b_{\phi}$ inference. The residuals in the recovered $\Delta b_{\phi}$ are reported as a function of the true $\Delta b_{\phi}$ (top left panel) and  as a function of the cumulative quantiles of old halos (top right panel). The different lines refer to  the different samples considered in the forecast analysis of Fig.~\ref{fig:sigmafnl_single}. The propagation of the residuals to the $\fnl$ fractional error is reported in the bottom left panel, while the bottom right panel shows the ratio of systematic to statistical errors for the LRG-like sample, for some representative values of $\fnl$. For small $\fnl$ values the systematic error can be safely kept  below  the statistical one.  The systematic error adopted here arises from taking the full difference $\Delta b_{\phi}^{(r)}-\Delta b_{\phi}^{(t)}$ without attempting to model the residuals (see text for more details).
}
\label{fig:sigma_fnl_sys}
\end{figure}

In the top left panel of Fig.~\ref{fig:sigma_fnl_sys}, we report the residuals as a function of the true $\Delta b_{\phi}$, for the relevant samples considered in this Section. As shown, the recovered $\Delta b_{\phi}^{(r)}$ of younger halos is overestimated, while that of older halos is underestimated. In the top right panel, the residuals are shown as a function of the cumulative proxy-selected $\omega_f$ quantiles of older halos. The behaviour of the residuals is strikingly similar for all the cases considered: ELG-like, LRG-like and full samples, also when for the latter one the proxy is calibrated on CAMELS instead of IllustrisTNG. This indicates that the residuals could be modeled as to reduce the systematic error on $f_{\rm NL}$ quite significantly. As an initial estimate of the maximum systematic contribution we assume it will not be corrected at all, and propagate the full extent of $\Delta b_{\phi}^{(r)}-\Delta b_{\phi}^{(t)}$. This is therefore a conservative estimate.

In the bottom left panel of Fig.~\ref{fig:sigma_fnl_sys} we report the relative (conservative) systematic error $\sigma^{(sys.)}_{f_{\rm NL}} \fnl^{-1}$, varying with the cumulative proxy-selected quantile of old halos, for all the samples considered in this Section. 

Including increasingly younger halos in the cumulative subsamples decreases the mean $\Delta b_{\phi}$ and the associated residual, as shown in the top panels. Consequently, this is mirrored by the systematic relative error in $\fnl$, which decreases with increasing size of the cumulative quantile, as illustrated in the bottom left panel. The performance of the different samples is similar to the results in Fig.~\ref{fig:sigmafnl_single}: the ELG-like sample has an overall larger error than the LRG-like one, while the full sample results lay in between the two. Again, the difference between the CAMELS- and the IllustrisTNG-calibrated proxies applied to the full sample (defined in Section~\ref{sec:proxy}) is marginal.

The relative systematic error associated to the LRG-like sample is the smallest mainly due to the $b_{\phi}$ larger than the ELG-like one. The relative contribution of the two types of errors depends on the subsample size considered, as well as on the value of $\fnl$: while the dependence of $\sigma_{\fnl}$ on $\fnl$ is weak, $\sigma^{(sys.)}_{f_{\rm NL}}$ scales linearly with it. Therefore, we expect the statistical error to be dominant for small $\fnl$, while the systematic one to dominate for large $\fnl$. This is illustrated in the bottom right panel of Fig.~\ref{fig:sigma_fnl_sys}, where we show the systematic to statistical relative error for the LRG-like sample, for several representative values of $\fnl$. For the quantile where the statistical error is the smallest ($\simeq 20-30$\%), the systematic error is below the statistical one as long as $\fnl \lesssim 10$.

\subsection{Multi-tracer}
By selecting subsamples with different clustering properties from the full sample of galaxies, one can employ the multi-tracer approach \cite{seljak_measuring_2009} and enhance the precision on $\fnl$ constraints. The improvement is due to the suppression of cosmic variance, which has a strong contribution on large scales, where we have most of the local PNG signal.
In this work we consider a 2-tracer approach (labeled hereafter $A$ and $B$), in which case the data vector is composed by the autopower spectra of the two tracers and the cross power spectrum between them: ${\bf D}(k)=\left\{P^A_{gg}(k),P^{A\times B}_{gg}(k),P^B_{gg}(k)\right\}$. By using Eq.~\eqref{eq:fisher}, one can obtain \cite{karagiannis_multi-tracer_2023} 
\begin{align}\label{eq:fisher_multi}
{\mathcal F}_{\fnl}&= \sum_{A,B}\int {\rm d}k \, {\mathcal N}^{-1}(k)\,\bigg\{ {4\,P_{\rm lin}(k)^2(b_{\phi}^B)^2\,\big[P_{\epsilon}^A+P_{\epsilon}^B\,R(k)^2\big]^2\,\left[b^{B}(k)\right]^2}\nonumber \\
&+{8\,P_{\rm lin}(k)^2\,P_{\epsilon}^B\,b_\phi^B\,\left[b_1^B\,b_\phi^A-b_1^A\,b_\phi^B\right]\,\big[P_{\epsilon}^A+P_{\epsilon}^B\,R(k)^2\big]\,R(k)\,b^B(k)} \nonumber \\
&+{2\, P_{\rm lin}(k)^2\left[b_1^B\,b_\phi^A-b_1^A\,b_\phi^B\right]^2\,\big[P_{\epsilon}^A\,P^{B}(k)+P_{\epsilon}^B\,\big\{ P_{\epsilon}^B+P^{B}(k)\big\}\,R^2(k)}\big] \bigg\}\;,
\end{align}
where
\begin{align}
{\mathcal N}(k)&\equiv \frac{2(2\pi)^3}{V_SV_{k}}\,{\cal M}(k,z_o)^2\big\{P_{\epsilon}^A\, P^{B}(k)+P_{\epsilon}^B \,\big[ P^{B}(k)-P_{\epsilon}^B \big]\, R(k)^2\big\}^2, \\
R(k)&\equiv \frac{b^A(k)}{b^B(k)}=\frac{b_1^A + b_{\phi}^A f_{\rm NL} {\cal M}^{-1}(k,z_o)}{b_1^B + b_{\phi}^B f_{\rm NL} {\cal M}^{-1}(k,z_o)}\;.
\end{align}

The key factor in this expression, encoding the dependence on the different bias parameters of the two traces is $|b_1^B\,b_\phi^A-b_1^A\,b_\phi^B|$ \cite{barreira_towards_2023,karagiannis_multi-tracer_2023}, which can be maximized by selecting subsamples with different specific properties in order to minimize $\sigma_{\fnl}$. 
The proxy developed in Sec.~\ref{sec:proxy}--\ref{sec:camels} can be used to select  such suitable subsamples.  Although the proxy is imperfect, as we show below,  it performs  very efficiently. 

\begin{figure}[t]
    \begin{minipage}{.32\textwidth}
	\includegraphics[width=1\linewidth]{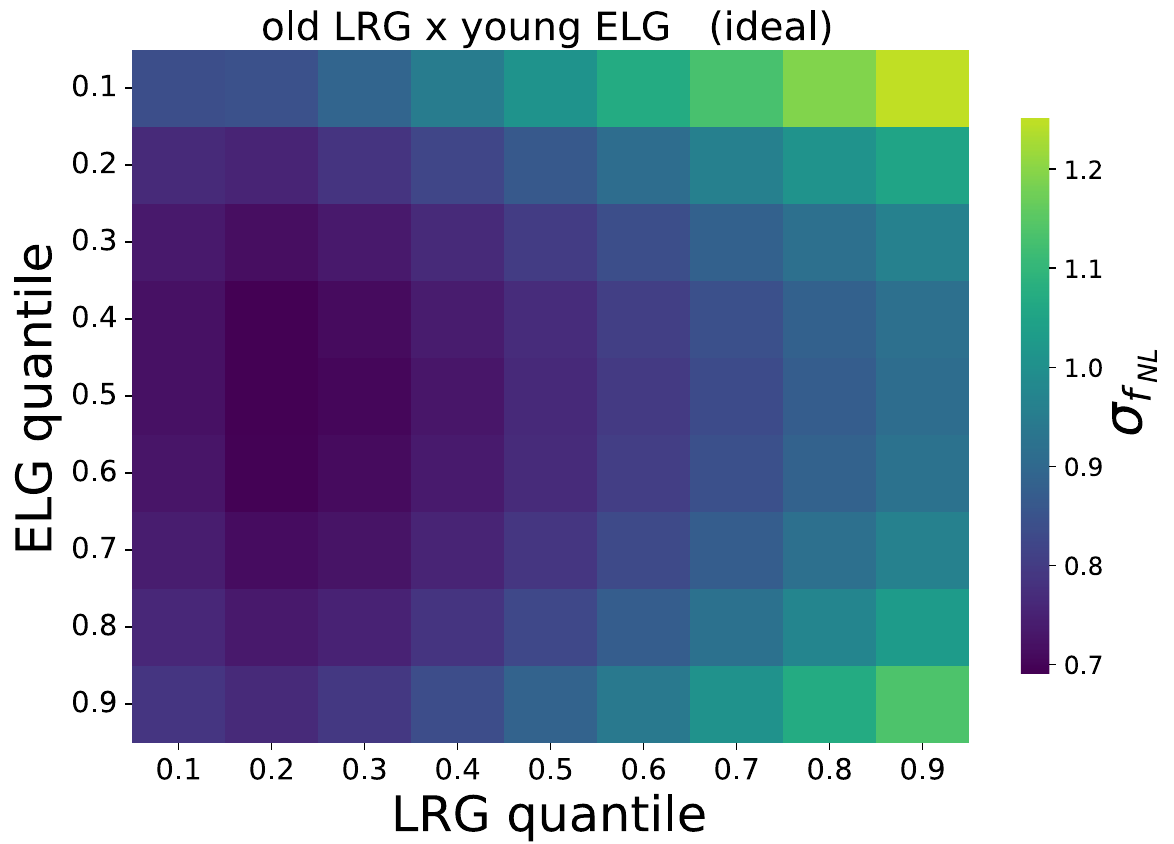}
    \end{minipage}
    \begin{minipage}{.33\textwidth}
	\includegraphics[width=1\linewidth]{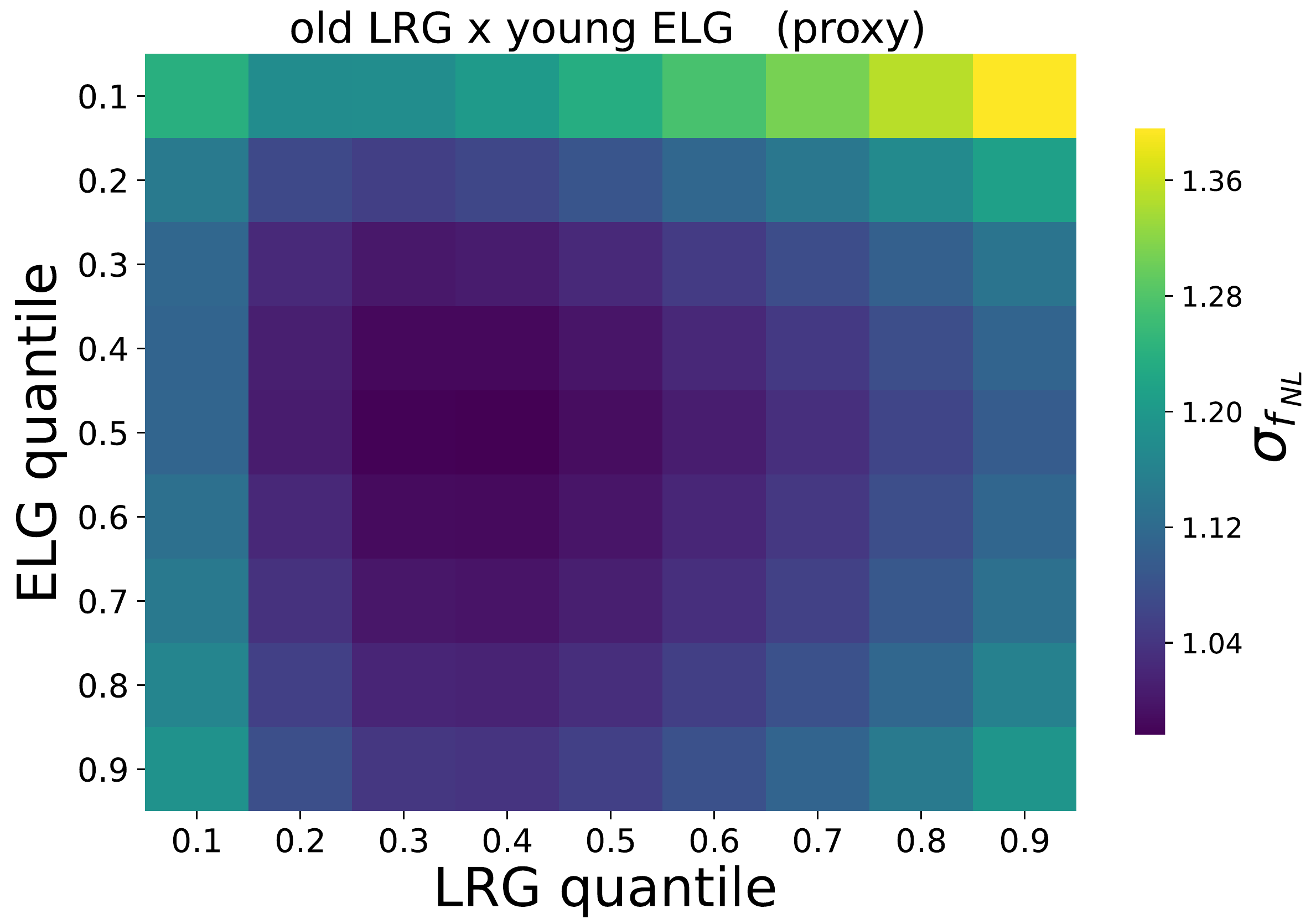}
    \end{minipage}
    \begin{minipage}{.33\textwidth}
	\includegraphics[width=1\linewidth]{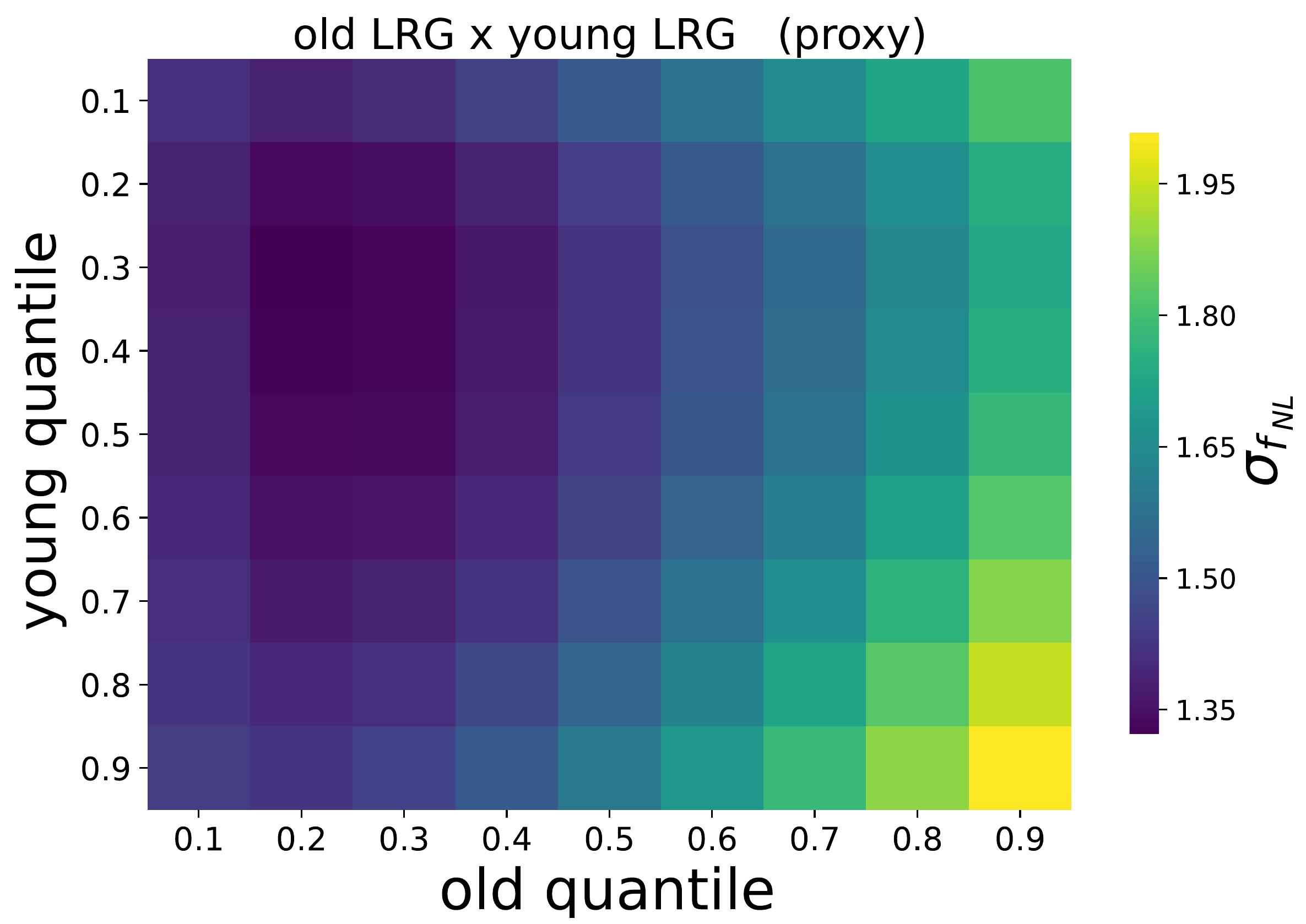} 
    \end{minipage}

    \caption{Statistical error $\sigma_{\fnl}$ in the multi-tracer case. For the two samples ELG- and LRG-like, we consider the combination of cumulative quantiles of the oldest and youngest halos. The results are reported as a function of the size of the cumulative subsamples, with color coding depending on the value of $\sigma_{\fnl}$. On the left, we show the results of the ideal case in which $\omega_f$ is known. Results relative to proxy-selected subsamples are shown in the middle and on the right.}
	\label{fig:sigmafnl_multi}
\end{figure}

In particular,  we use the proxy to select and  combine large $b_\phi^A$  with small $b_\phi^B$ subsamples, as to maximize the mentioned factor. In Fig.~\ref{fig:sigmafnl_multi} we report the results on $\sigma_{\fnl}$,  as a function of  the size of ELG- and LRG-like proxy-selected cumulative quantiles.
The numerical values used for the computation of Eq.~\eqref{eq:fisher_multi} are the same as used in the single-tracer case. We refer to the subsample of LRG-like galaxies hosted by old halos as ``old LRG" for brevity, and similarly for ``young ELG". We do not consider ``old ELG" subsamples as their $\Delta b_{\phi}^{(r)}$ is smaller than in the LRG case.

The results in Fig.~\ref{fig:sigmafnl_multi} illustrate the power of the multi-tracer technique. If we consider the LRG-like sample, as shown in Fig.~\ref{fig:sigmafnl_single}, the error on $\fnl$ can be brought from $\sigma_{\fnl}\gtrsim 4$ to $\sigma_{\fnl}=3.2$ by selecting a subsample of galaxies hosted by old halos, a marginal improvement. However, if this subsample is cross-correlated  (via the multitracer approach) with a selection of ELGs formed in young halos, as shown in Fig.~\ref{fig:sigmafnl_multi}, we can get an improvement of a factor of 4 in the statistical errors,   obtaining $\sigma_{\fnl}\simeq1$. Although this forecast may be further improved by adopting a better proxy, it is only $\simeq30$ \% larger than the ideal case in which $\omega_f$ is known, as shown in the left panel of Fig.~\ref{fig:sigmafnl_multi}. Due to the broader $\omega_f$ distribution of the ELG-like sample, a subsample of young halos can be selected efficiently, while this does not apply for the LRG-like sample. Therefore, cross-correlating old and young halos hosting LRGs lead to suboptimal results, as illustrated in the right panel of Fig.~\ref{fig:sigmafnl_multi}.

In both cases, similarly to the single-tracer results, we can identify the optimal fractions which minimize $\sigma_{\fnl}$, roughly corresponding to selecting a third of the full sample. As already discussed in Section 5.1, this is related to the tradeoff between signal and (shot) noise driven by the number density.

As regards the systematic error on $\fnl$, induced by systematic shifts in the estimated $b_{\phi}$ of the two multi-tracer samples, we refer to \cite{barreira_towards_2023} and evaluate it by using the following expression
\begin{equation}
\frac{\sigma^{(sys.)}_{\fnl}}{\fnl} = \frac{b_{\phi}^{A, t}\,b_{\phi}^{A, r}/\sigma_A^2 \; + \; b_{\phi}^{B, t}\,b_{\phi}^{B, r}/\sigma_B^2}{\left(b_{\phi}^{A, r}\right)^2/\sigma_A^2 \; + \; \left(b_{\phi}^{B, r}\right)^2/\sigma_B^2}-1
\label{eq:sys_shift_multi}
\end{equation}
where the subscript \textit{t}(\textit{r}) indicates the true (recovered) $b_{\phi}$ of the sample and $\sigma_{A,B}$ stands for the single tracer's statistical error on $\fnl$.

Similarly to the single tracer case, the systematic error scales linearly with $\fnl$. Differently from the previous case, as discussed in \cite{barreira_towards_2023}, the multi-tracer technique should be more robust to systematic shifts in $b_{\phi}$. However, Ref. \cite{barreira_towards_2023} assumed both $b_{\phi}$ of the two samples to be shifted in the same direction, by the same amount, in which case their effects in Eq.~\eqref{eq:sys_shift_multi} cancel each other. In our case, instead, the residuals of the sample containing young halos have opposite sign with respect to those related to old halos, as shown in the top left panel of Fig.~\ref{fig:sigma_fnl_sys}. Therefore, this cancellation does not apply here.

\begin{figure}[H]
\centering
    \begin{minipage}{.45\textwidth}
	\includegraphics[width=1\linewidth]{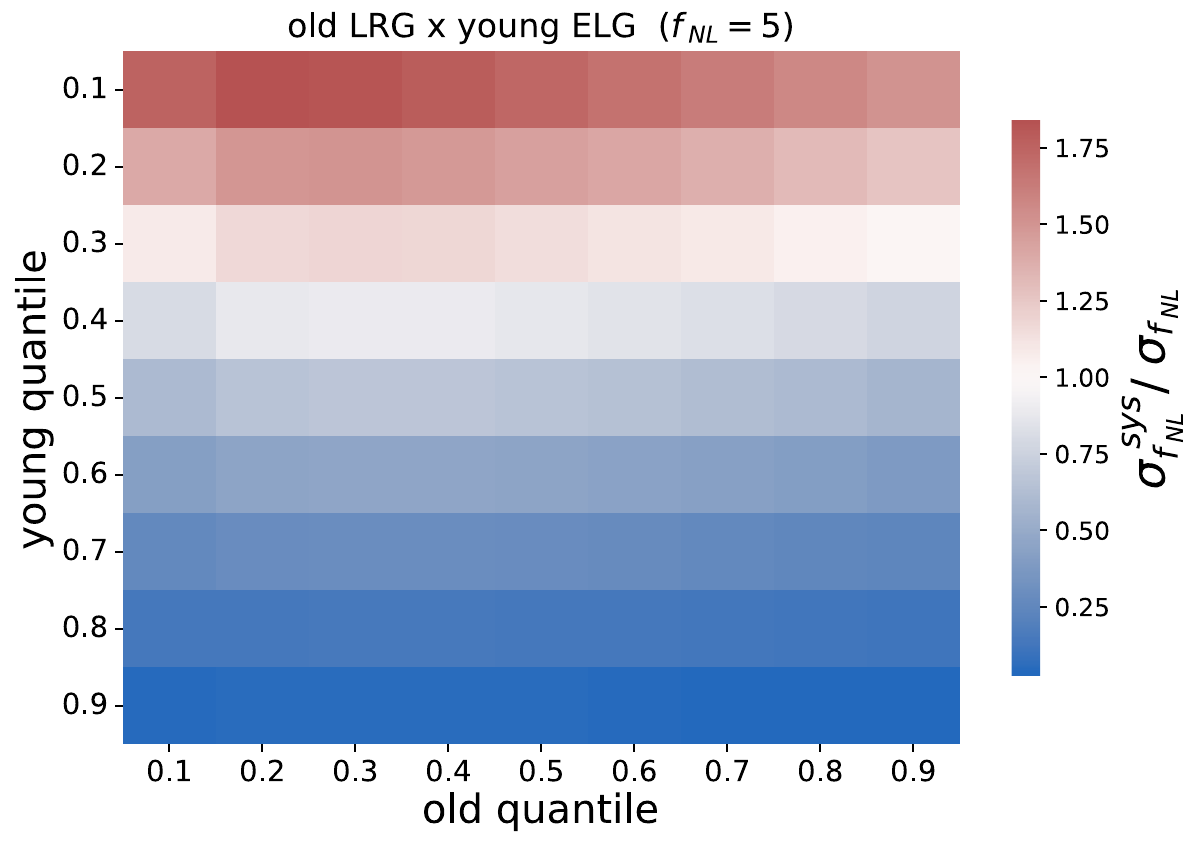}
    \end{minipage}
    \hspace{0.8cm}
    \begin{minipage}{.45\textwidth}
	\includegraphics[width=1\linewidth]{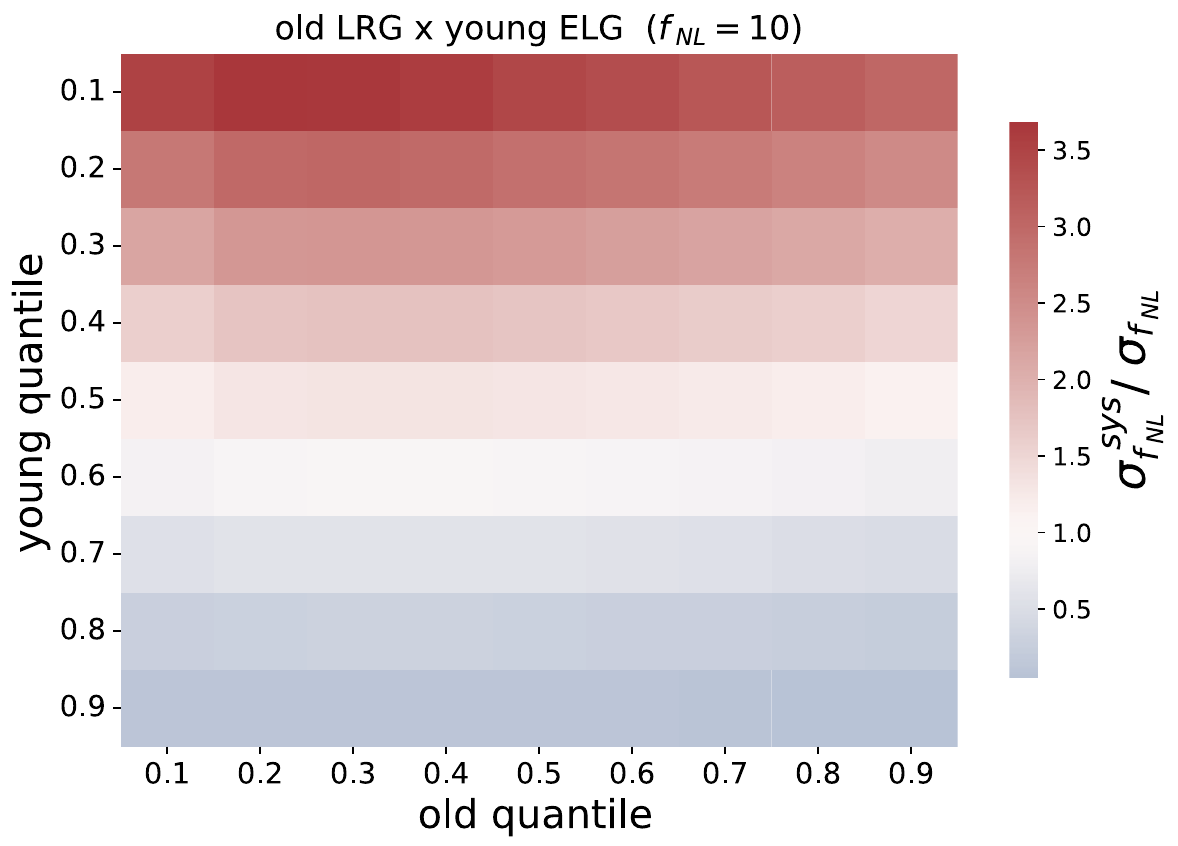} 
    \end{minipage}
	\caption{$\sigma_{\fnl}^{(sys.)}$ in the multi-tracer case. The samples considered are the old LRG-like (large $b_{\phi}$) and the young ELG-like (small $b_{\phi}$) ones. The different panels illustrate the results for the representative values of $\fnl=5,10$, as a function of the size of the cumulative subsamples. In all the plots, the white color denotes the limit in which the systematic error is larger than the statistical one.}
	\label{fig:sigmafnl_sys_multi}
\end{figure}

In Fig.~\ref{fig:sigmafnl_sys_multi} we show the ratio between the systematic error in $\fnl$ and the statistical one, as a function of the quantiles of the two samples, for different representative values of $\fnl$.

In the same way as illustrated in Fig.~\ref{fig:sigma_fnl_sys}, as expected due to the linear scaling with $\fnl$ of the systematic error, for values as low as $\fnl\lesssim5$, the systematic error is a fraction of the statistical one. For larger values of $\fnl$, the systematic error can be reduced to be below the statistical one by choosing a larger subsample size, for which the $b_{\phi}$ systematic shift is lower. Although this may be a suboptimal choice from the statistical error point of view, it is worth to highlight that if taken at face value, the multi-tracer systematic error is significantly lower than the single-tracer one.

\begin{figure}[H]
\centering
\includegraphics[width=0.8\linewidth]{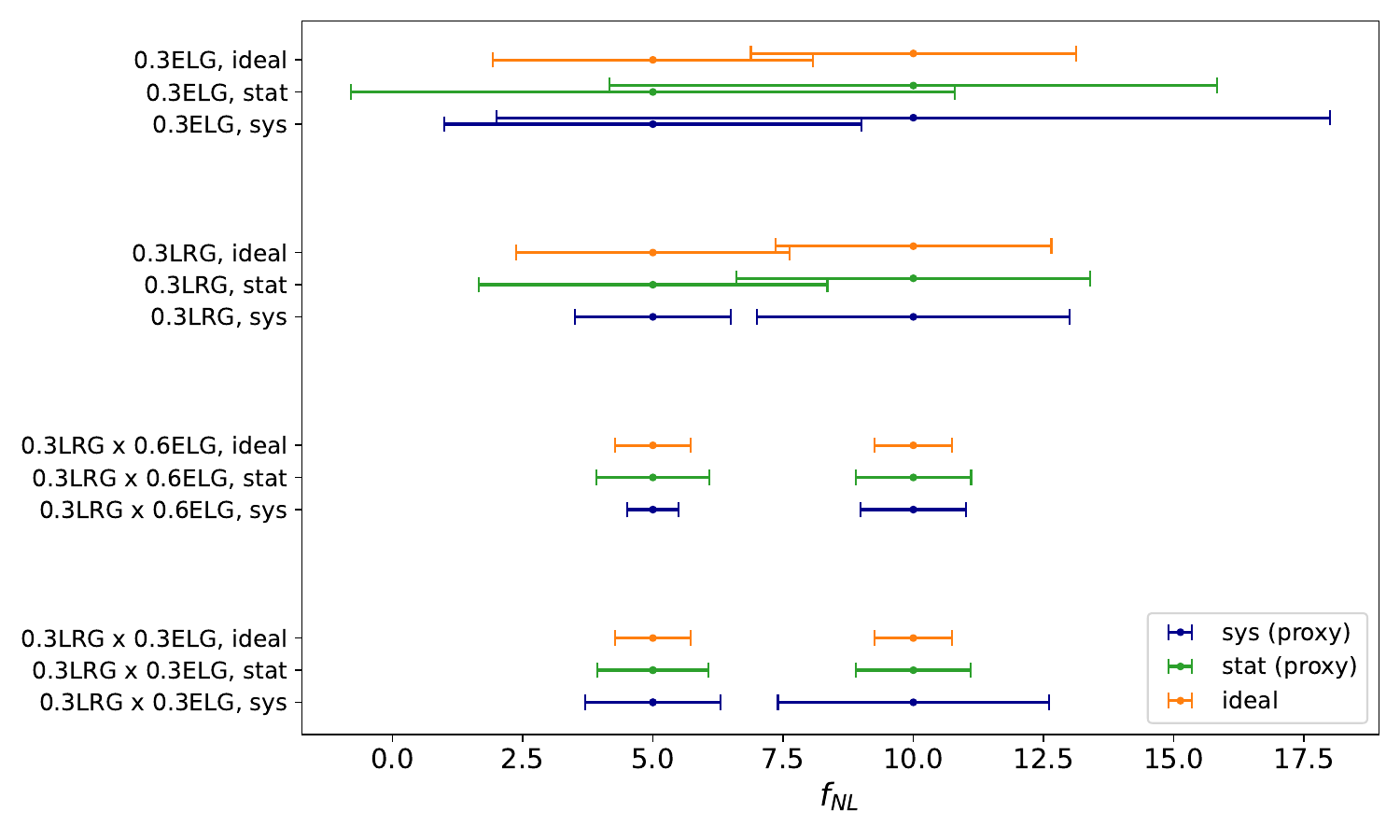}

\caption{Summary of the main $\fnl$ error forecast results reported throughout Section \ref{sec:forecasts}. For each of the two representatives values $\fnl=5,10$, we show the forecasted error for the ideal and the proxy-selected quantiles, reporting both the statistical and the systematic error in the latter case. For the single tracer case (top  two groups), results are shown for the 30 \% quantile of oldest halos in the LRG and ELG-like samples. As for multi-tracer, we consider the ``old LRG X young ELG" combination, with two different choices of quantile sizes. We recall that statistical errors are for a volume of $20.4$ Gpc$^3/h^3$.}
\label{fig:sigmafnl_summary}
\end{figure}

Fig.~\ref{fig:sigmafnl_summary} is an attempt to  summarize the main takeaways of this Section. We report a selection of forecasts for  two different representative values $\fnl=5,10$. We recall that statistical errors are for a volume of $20.4$ Gpc$^3/h^3$.

We also reiterate that the systematic error estimate  is also likely conservative, as we have taken the full residuals into account without modeling them. 

\section{Conclusions}
\label{sec:conclusions}
The presence of local Primordial non-Gaussianity (PNG) in the initial conditions of the Universe  affects dark matter clustering, inducing a scale dependence in the halo bias on large scales. This feature provides, in principle, a very competitive approach for constraining the amplitude of local PNG, parameterized by  $\fnl$. However, the determination of $\fnl$ critically depends on the  knowledge of the local PNG bias parameter $b_{\phi}$, currently the subject of ongoing discussions due to its sensitivity to assembly bias. Ref.~\cite{reid_non-gaussian_2010} investigated the non-Gaussian halo assembly bias, explicitly showing the dependence of $b_{\phi}$ on halo formation time and providing theoretical predictions based on the extended Press-Schechter prescription. Our work serves as an extension of this prior research, highlighting how assembly bias can be modeled very accurately from first principles and priors on $b_{\phi}$ can be imposed by considering the conditional halo mass function. This is confirmed by tests on N-body simulations. 

While theoretical predictions apply to dark matter halos, observations provide galaxy surveys. The poorly known halo-galaxy connection represents a limitation to the application of the theoretical modeling of $b_{\phi}$ to data and thus to obtain observational constraints on $\fnl$. 
This work provides a first attempt to bridge this gap.

We propose an observational proxy whereby  photometric properties of galaxies can be used  to infer  statistical information on  the assembly history of their host halos. The proxy can then be used to identify assembly-bias-selected  subsamples and  provide an estimate of  the subsamples'  $b_{\phi}$. In doing so   we also identify which types of galaxy samples represent optimal targets to provide more precise constraints on $\fnl$.

Our main results can be summarized as follows:

\begin{itemize}
    \item At fixed mass, halos that accreted their mass earlier have a larger $b_{\phi}$ than predicted by the universality relation and viceversa for late-forming halos. The departure from universality can be related to the conditional halo mass function, independently of its specific functional form (this effect goes under the name of assembly bias).
    \item The ePS formalism allows us  to derive an analytical prediction for the conditional mass function, which is independent of halo mass and redshift. However, its accuracy decreases when considering halos of mass $M\lesssim 10^{13} M_{\odot} h^{-1}$, while a simple 1-parameter extension shows a remarkably good fit to simulations across all relevant mass and redshift ranges.
    Hence the assembly bias and thus $b_{\phi}$ for a given halo sample depends through a simple  (physically motivated and calibrated on N-body simulations) analytic function  on a single variable $\omega_f$,  which encodes a specific property of the halo sample assembly history.
    
    \item  Using state-of-the art  cosmological simulations that model both the dark matter clustering and  galaxy properties (IllustrisTNG and the CAMELS-TNG suite) we calibrated an observational proxy for $\omega_f$, the quantity  which determines halo assembly bias and thus $b_{\phi}$. The proxy is built (by design) from a linear combination of photometric bands (i.e. colors) of the galaxies and tuned to  minimize the difference between the true $b_{\phi}$ of a given halo sample and the one recovered from the photometry of the  halos' central galaxies.

    \item The proxy is used to identify subsamples  which are (statistically)  rank ordered according to their $\omega_f$ and thus  $b_{\phi}$ and to provide an estimate of their $b_{\phi}$. The proxy is of course not perfect: $\omega_f$ is recovered with some scatter and $b_{\phi}$ is recovered with some dispersion and a systematic shift which depends on the sample selection.   

    \item Marginalization over cosmological and IllustrisTNG astrophysical parameters (performed with the CAMELS-TNG simulations suite) does not degrade the proxy performance significantly. However, further work should be done to assess the robustness on galaxy formation models beyond IllustrisTNG.

    \item The  statistical error on the  $b_{\phi}$ recovered from the proxy  does not degrade significantly the forecasted errors on $\fnl$ compared to the ideal case when $b_{\phi}$ is perfectly known.
   
    \item The systematic error on  the proxy-recovered $b_{\phi}$ is remarkably regular across different galaxy selection strategies. This indicates that the shift could be modeled and subtracted. The proxy performance could  probably be improved by generalizing it to a non-linear combination of galaxies' colors. Even without these improvements, and taking the full uncorrected  shift in $b_{\phi}$ as a source of systematic error on $\fnl$, we show that the resulting  systematic error on $\fnl$ can be kept reasonably under control.
    
    \item The proxy can also be used to  observationally select halo subsamples with $b_{\phi}$ as different as possible as to optimize the constraints on $\fnl$ with the  multi-tracer approach. In this case, forecasted errors on $\fnl$ reach the level of $\pm \mathscr{O}(1)$ for volumes $\sim \mathscr{O} (20)$ Gpc$^3$.   
\end{itemize}
We hope that the findings of this work  will serve to demystify the  impact of assembly bias on PNG constraints and will motivate improvements on the  simple observational proxy  proposed here. Even just with the naive, linear combination of central galaxy colors proxy, the power spectrum of biased tracers offer a competitive and viable window into PNG.  

\acknowledgments
EF acknowledges the support from ``la Caixa” Foundation (ID 100010434, code LCF/BQ/DI21/11860061).
Funding for this work was partially provided by project PGC2018-098866-B-I00 \\ MCIN/AEI/10.13039/501100011033 y FEDER “Una manera de hacer Europa”, and the “Center of Excellence Maria de Maeztu 2020-2023” award to the ICCUB (CEX2019-000918-M funded by MCIN/AEI/10.13039/501100011033).
LV acknowledges support of  European Union's Horizon 2020 research and innovation programme ERC (BePreSysE, grant agreement 725327).
The CAMELS project is supported by the Simons Foundation and NSF grant AST 2108078.
AR acknowledges support from PRIN-MIUR~2020 METE, under contract no. 2020KB33TP. DK is supported by the South African Radio Astronomy Observatory and the National Research Foundation (Grant No.\ 75415). GJ acknowledges support from the ANR LOCALIZATION project,
grant ANR-21-CE31-0019 / 490702358 of the French Agence Nationale de la Recherche. The Center for Computational Astrophysics and the  Flatiron Institute are supported by the Simons Foundation.

\appendix

\section{Appendix}
\label{appendix}
\begin{figure}[H]
\centering
\includegraphics[width=0.6\linewidth]{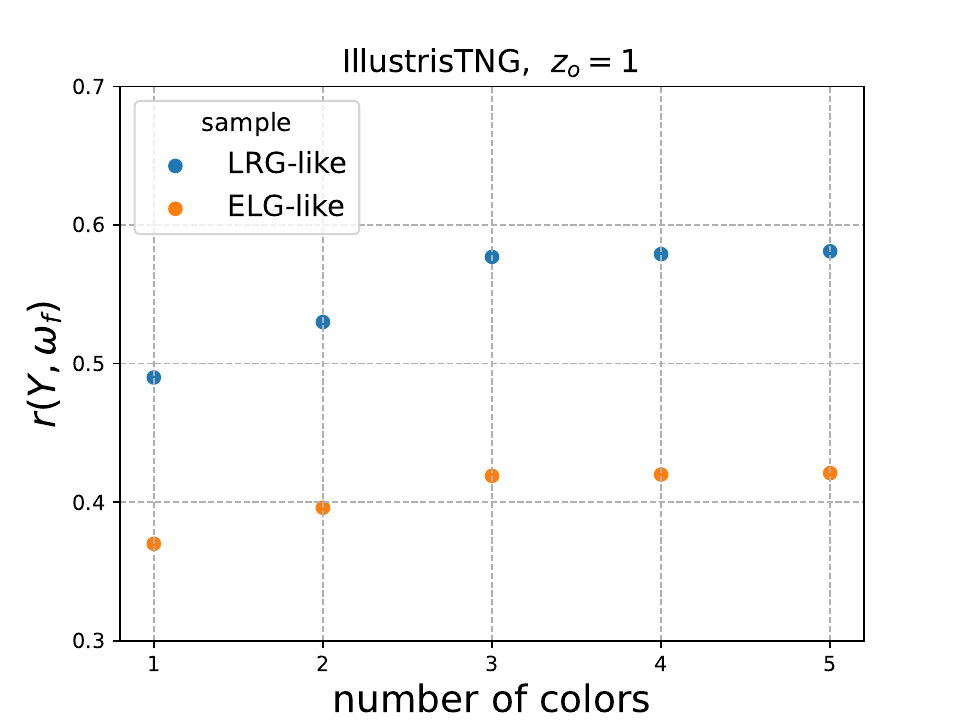}

\caption{Pearson correlation coefficient $r(Y,\omega_f)$ between the observational proxy $Y$ and $\omega_f$, as a function of the number of colors included in the linear combination which defines $Y$. The results are reported for the two representative samples of ELG- and LRG-like galaxies in the IllustrisTNG simulations, at $z_o=1$.}
\label{fig:ncolors_corr}
\end{figure}

As discussed in Sec.~\ref{sec:proxy}, in order to estimate $\omega_f$ from observational galaxy properties, we make use of the information contained in IllustrisTNG galaxy colors. In this Appendix we show that, as long as a simple linear combination $Y=\sum\limits_{i=1}^N a_i C_i$ is considered, $N=3$ colors are  sufficient to maximize the Pearson correlation $r(Y,\omega_f)$. In particular, in Fig.~\ref{fig:ncolors_corr} we report $r$ as a function of the number of colors $N$ for two representative galaxy samples, the ELG- and LRG-like in IllustrisTNG at $z_o=1$.

As illustrated in Fig.~\ref{fig:ncolors_corr}, adding colors to the combination improves the correlation between the optimal proxy Y and $\omega_f$, until $N=3$. Then $r$ saturates and the addition of further colors becomes redundant. Specifically, for the LRG-like sample, $r$ only increases by $0.6$\% passing from a combination of 3 colors to a combination of 5, while the computational time needed to optimize the model is 30 times larger. Similar results apply for what regards the ELG-like sample.

Consequently, for all the galaxy samples considered in Sec.~\ref{sec:proxy}, we restrict our optimization of the proxy to a linear combination of 3 galaxy colors.
\bibliographystyle{JHEP}

\providecommand{\href}[2]{#2}\begingroup\raggedright\endgroup

\end{document}